\documentclass[smallextended]{svjour3}
\usepackage[utf8]{inputenc}
\usepackage[dvipsnames]{xcolor}
\definecolor{BlizzardBlue}{rgb}{0.67, 0.9, 0.93}
\usepackage{graphicx}
\usepackage{float,wrapfig}
\usepackage[hidelinks,breaklinks=true]{hyperref}
\usepackage[nocompress]{cite}
\usepackage{url}
\graphicspath{{img/}}
\usepackage{footmisc}
\usepackage{booktabs,tabularx,multirow,pdflscape}
\usepackage[toc,page]{appendix}
\usepackage{caption}
\usepackage{subcaption}
\usepackage{setspace}
\usepackage{enumitem}

\newlist{myQuoteEnumerate}{enumerate}{2}
\setlist[myQuoteEnumerate,1]{label=(\arabic*)}
\setlist[myQuoteEnumerate,2]{label=(\alph*)}

\title{On the evolution and impact of Architectural Smells - An industrial case study}

\author{ Darius Sas \and Paris Avgeriou \and Umut Uyumaz }
\institute{Darius Sas, Paris Avgeriou 
    \at Bernoulli Institute for Mathematics, Computer Science, and Artificial Intelligence, University of Groningen, Bernoulliborg Nijenborgh 9, 9747 AG Groningen, Netherlands\\\email{\{d.d.sas,p.avgeriou\}@rug.nl}
    \\ORCID: 0000-0003-3383-3298, 0000-0002-7101-0754
\and Umut Uyumaz \at ASML, De Run 6501, 5504 DR Veldhoven, Netherlands\\\email{umut.uyumaz@asml.com}}
\date{\today}
\authorrunning{Sas et al.}
\begin{document}
\maketitle
\begin{abstract}
Architectural smells (AS) are notorious for their  long-term impact on the Maintainability and Evolvability of software systems.
The majority of research work has investigated this topic by mining software repositories of open source Java systems, making it hard to generalise and apply them to an industrial context and other programming languages.
To address this research gap, we conducted an embedded multiple-case case study, in collaboration with a large industry partner, to study how AS evolve in industrial embedded systems.
We detect and track AS in 9 C/C++ projects with over 30 releases for each project that span over two years of development, with over 20 millions lines of code in the last release only.
In addition to these quantitative results, we also interview 12 among the developers and architects working on these projects, collecting over six hours of qualitative data about the usefulness of AS analysis and the issues they experienced while maintaining and evolving artefacts affected by AS.
Our quantitative findings show how individual smell instances evolve over time, how long they typically survive within the system, how they overlap with instances of other smell types, and finally what the introduction order of smell types is when they overlap.
Our qualitative findings, instead, provide insights on the effects of AS on the long-term maintainability and evolvability of the system, supported by several excerpts from our interviews. Practitioners also mention what parts of the AS analysis actually provide actionable insights that they can use to plan refactoring activities.
\end{abstract}

\section{Introduction}
Architectural decisions have been established as one of the most important factors affecting long-term maintenance and evolution of software systems \cite{Ernst2015}.
Architectural smells (AS) are a specific type of such decisions; they are defined by Garcia et al. as \emph{``commonly-used (although not always intentional) architectural decisions that negatively impact system quality''}  \cite{Garcia2009}.
There are several research works that define the different \emph{types} of architectural smells (e.g. god components or cycles between components) and discuss their impact on maintainability and other qualities \cite{Lippert2006, Fontana2016, Mo2015, Le2016, Garcia2009}.
This impact usually depends on the type of smell, but generally, an architectural smell can impact maintenance activities of all kinds (corrective, perfective, etc.) by violating software design principles \cite{Azadi2019}.
For example, AS can hinder the adaptation of a system to new requirements by increasing the coupling and breaking the modularity of certain parts of the system \cite{Azadi2019}.

Despite the significant corpus of research available on the topic \cite{Verdecchia2018}, most studies have a \textbf{limited scope} as they perform mainly \emph{source code} analyses on \emph{open source} systems written in \emph{Java}. 
While these studies certainly provide a valid and substantial contribution to the literature, there is insufficient work on real-world industrial systems. Particularly,  to the best of our knowledge, there is no work on the impact of AS on maintainability in the embedded systems (ES) industry,
where  languages like C, C++, and Python are used much more than Java \cite{Tiobe2021}\footnote{See our replication package for the version at the moment of writing.}.

To address this shortcoming, this study investigates AS in an \emph{industrial setting} by analysing \emph{C/C++} projects and eliciting the \emph{opinion} of software engineers and architects.
In particular, we worked with an industrial partner, ASML\footnote{\label{fn:asml}Visit \url{www.asml.com} for more info.}, and studied how \emph{AS evolve} and impact Maintenance and Evolution \cite{Vliet2008} in two steps.
First, we studied  the evolution of AS in one of ASML's main software product lines, comprised of several millions of lines of code, by examining: how architectural smell instances evolve in terms of their \textbf{characteristics} (e.g. number of affected elements, number of dependency edges among the affected elements, etc.), how long they \textbf{persist} in the system, and how they \textbf{overlap}.
Second, we showed the architects, designers, and developers the results of our analysis and \textbf{interviewed them about the issues they experience} while maintaining the artefacts affected by architectural smells.
This study design allowed us to cover the viewpoints of both the system (quantitative) and the engineers (qualitative). 

The major findings of this study show that smells tend to grow \emph{larger over time}, affecting more and more artefacts, and that different smell types exhibit largely different survival rates, allowing practitioners to do a coarse-grained prioritisation of the smells instances to refactor. 
Moreover, the results show that some artefacts are affected by more than one smell at a time, \emph{increasing} the effort required to maintain them.
Practitioners, on the other hand, recognise that the presence of smells correlates with frequently changed components, increased change propagation, the presence of severe bugs, the decay of the architecture, and general maintenance issues.

The architectural smells considered in this study are Cyclic Dependency (CD), Hub-Like Dependency (HL), Unstable Dependency (UD), and God Component (GC) \cite{Fontana2016,Lippert2006,Sas2019}. We opted to study these smells as they are some of the most prominent architecture smells, and there already exists tools that support their automatic detection \cite{Fontana2016, Fontana2017}.

The rest of the paper is organised as follows: Section \ref{sec:background} provides a brief overview on the architectural smells subject of this study as well as their characteristics and the tool used to detect them; Section \ref{sec:related-work} discusses related work and compares it with this study; Section \ref{sec:case-design} provides a detailed description of the study design; Sections \ref{sec:rq-1}, \ref{sec:rq2}, \ref{sec:rq-3}, and \ref{sec:rq-4-and-5} describe the data analysis methodology and results for each research question; Section \ref{sec:discussion} provides a discussion on the findings presented in the previous sections; Section \ref{sec:implications} summarises the implications of our findings for practitioners; Section \ref{sec:threats-to-validity} summarises the identified threats to the validity and our mitigation strategy; finally, Section \ref{sec:conclusion-fw} concludes the paper and lists some possibilities of future work.
Appendix \ref{appendix:interview-guide} reports the interview guide we used during the interviews.

\section{Background}\label{sec:background}
\subsection{Architectural smells: Definitions and implications}\label{sec:arch-smells}
This section lists the architectural smells (AS) considered by this study. The definition of these smells is provided by Arcelli et al. \cite{Fontana2016} and briefly reported here.

\begin{figure}[t]
    \centering
    \begin{subfigure}[t]{0.45\linewidth}
        \centering
        \includegraphics[width=0.5\linewidth]{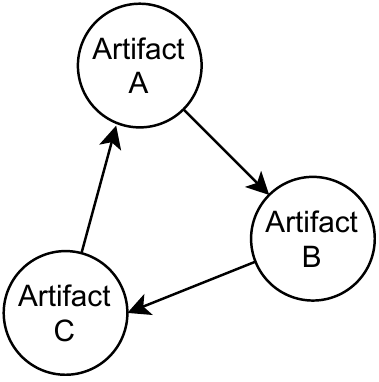}
        \caption{An example of Cyclic Dependency among artefacts A, B, C.}\label{fig:cd}
    \end{subfigure}
    \hfill
    \begin{subfigure}[t]{0.45\linewidth}
        \centering
        \includegraphics[width=0.7\linewidth]{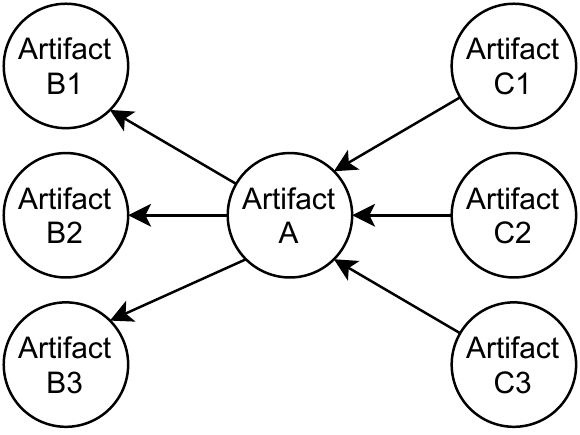}
        \caption{An example of HL affecting component A, with the afferent (incoming) dependencies on the right and (outgoing) efferent dependencies on the left.}\label{fig:hl}
    \end{subfigure}
    \\
    \begin{subfigure}[t]{0.45\linewidth}
        \centering
        \includegraphics[width=0.7\linewidth]{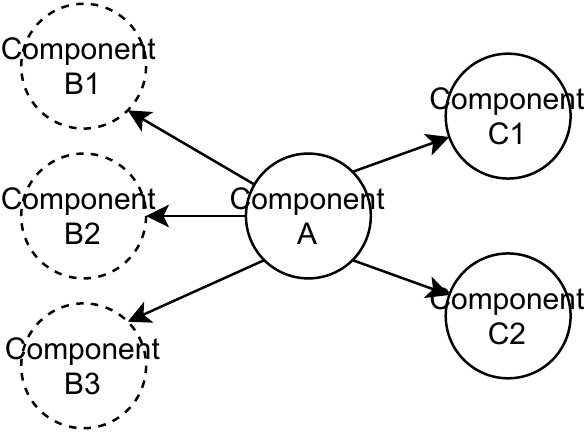}
        \caption{An example of UD affecting component A. The components that A depends on (Bs and Cs) are shown in the figure too. Components B1 to B3 are less stable than A and represent the majority of A's outgoing dependencies.}\label{fig:ud}
    \end{subfigure}
    \hfill
    \begin{subfigure}[t]{0.45\linewidth}
        \centering
        \includegraphics[width=0.5\linewidth]{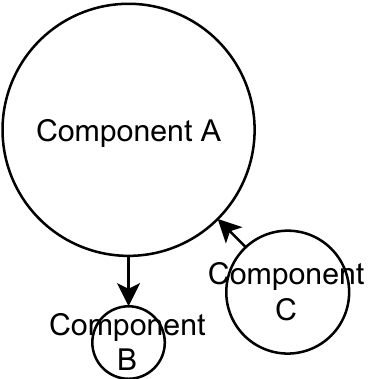}
        \caption{An example of GC affecting component A. The diameter of A represents its larger size in terms of lines of code w.r.t. other components in the system (B and C).}\label{fig:gc}
    \end{subfigure}
    \caption{Illustration of the four architectural smell types considered in this work.}
    \label{fig:architectural-smells} 
\end{figure}

\paragraph{Unstable dependency (UD)}\label{sec:arch-smells-ud}
This smell represents a component that depends upon a significant number of components that are less stable than itself.
The stability of a component is measured using Martin's instability metric \cite{martin1994}, which measures the degree to which a component (e.g. a package) is susceptible to change based on the classes it depends upon and on the classes depending on it.
The smell thus arises when a component has a significant number of components -- the tool \textsc{Arcan} uses a 30\% threshold \cite{Fontana2017} -- it depends upon with an instability value higher than its own.
A UD smell is detectable on Java package-like elements only (i.e. containers of classes, files, etc.). A simplified example of UD is shown in Figure \ref{fig:ud}. 

The main problem caused by UD is that the probability to change the main component grows higher as the number of unstable components it depends upon grows accordingly. This increases the likelihood that the components that depend upon it (not shown in Figure \ref{fig:ud} for simplicity) change as well when it is changed (ripple effect), thus inflating future maintenance efforts.

\paragraph{Hublike dependency (HL)}\label{sec:arch-smells-hl}
This smell represents a component where the number of ingoing and outgoing dependencies is higher than the median in the system and the absolute difference between these ingoing and outgoing dependencies is less than a quarter of the total number of dependencies of the component \cite{Fontana2016}. A hublike dependency can be detected both at the package and at the class level.

The implications of this smell for development activities are once again concerning the probability of change and the ease of maintenance. Consider, for example, the case represented in Figure \ref{fig:hl}.
Making a change to any of the components that A depends upon may be very hard \cite{martin1994}, even though there is only one component depending on them.
Additionally, the central component is also overloaded with responsibility and has a high coupling.
This structure is thus not desirable, as it increases the potential effort necessary to make changes to all of the elements involved in the smell.

\paragraph{Cyclic dependency (CD)}\label{sec:arch-smells-cd}
This smell represents a cycle among a number of components; there are several software design principles that suggest avoiding creating such cycles \cite{Lippert2006,Parnas1979,Stevens1974,Martin2000}.
Cycles may have different topological shapes. Al-Mutawa et al. \cite{AlMutawa2014} have identified 7 of them; the ones detected by \textsc{Arcan} are shown in Figure \ref{fig:cycle-shapes} \cite{Fontana2017}.
Usually, the circle shape is intuitively perceived as the typical CD (i.e. see Figure \ref{fig:cd}), but it is certainly not the only possible type of CD. In fact, there is empirical evidence \cite{AlMutawa2014} that tiny and multi-hub shapes (two stars attached together that are missing some edges) are more common than one expects.
More complex shapes mean that the cycle has lower levels of coupling and higher levels of cohesion among the elements creating the cycle.
For example, a clique-shaped cycle has the maximum amount of coupling possible with the components taking part in the cycle, drastically reducing the maintainability of the affected part of the system.

Besides affecting complexity, their presence also has an impact on compiling (causing the recompilation of big parts of the system), testing (forcing to execute unrelated parts of the system, increasing testing complexity), or deploying (forcing developers to re-deploy unchanged components) \cite{Lippert2006}.

\paragraph{God component (GC)}\label{sec:arch-smells-gc}
This smell represents a component (or package, in Java) that is considerably larger in size (i.e. lines of code) than other components in the system \cite{Lippert2006} (see Figure \ref{fig:gc}).
Originally, GC was defined using a fixed threshold on the lines of code, \textsc{Arcan} however uses a variable threshold-detection approach based on the frequencies of the number of lines of code of the other packages in the system \cite{Fontana2015}.

God components aggregate too many concerns together in a single artefact and they are generally a sign that there is a missing opportunity for splitting up the component into multiple sub-components.
God components tend to become such over time, as a result of several little incremental changes that contribute to the massive scale of the component, which ends up effectively implementing a lot of the overall functionality of the system.
Over time, the understandability of the component deteriorates along with the reusability of the individual parts of the component, because nobody wants to use a piece of software that is difficult to understand \cite{Lippert2006}.

\subsection{Architectural smell characteristics}\label{sec:characteristics}
An architectural smell \emph{characteristic} is a property or attribute of an architectural smell instance \cite{Sas2019}. 
An architectural smell \emph{instance} is a concrete occurrence of a type of architectural smell.
Smell instances can span over multiple consecutive versions; in that case we refer to them as \emph{temporal instances}.
For each architectural smell type, one can measure different characteristics. Some characteristics can be measured for every type of smell; we refer to them as \emph{smell-generic}. Other characteristics can only be measured for certain types of smells; we refer to those as \emph{smell-specific} characteristics.
The characteristics considered in this work are reported in Table \ref{tab:characteristics}.

We opted to focus our analysis on this set of smell characteristics because they provide further insights about the extent that a smell affects the system; this can inform developers on how to prioritise refactoring.
Additionally, some of the selected characteristics were developed, studied or discussed by other authors in previous studies, as denoted in Table \ref{tab:characteristics}.

The smell-generic characteristics \emph{Centrality}, \emph{Size}, and \emph{Number of edges} are of interest because they are all conceptually related to the complexity caused by an instance of a smell in the system. 
Intuitively, smells that affect parts of the system that are more central (centrality) in the dependency network of the system are likely to also affect parts that are critical and complex, and therefore harder to refactor and maintain. 
Centrality is measured using PageRank \cite{Page1999}, an algorithm that measures the relative importance of a node in a given network based on the connections among the nodes themselves. The more nodes point to a node, and the more important those nodes are, the more importance is assigned to that node.
PageRank was used on software dependency networks in a previous study \cite{Roveda2018b}.

Smells that affect more elements (i.e. larger \emph{Size}) have a greater impact on the system's maintainability and are more complex to comprehend.
Likewise, smells that affect elements that are highly connected among them (number of edges), are more complex (because they correspond to higher coupling) and therefore have higher impact on maintainability.

The \emph{Age} of a smell instance keeps track of the number of consecutive versions the smell was present in.
Architects and developers can decide to prioritise refactoring based on the age of a smell instance. For example, one might want to focus on the newly introduced smells to ensure the new code added to the system is more maintainable.

The CD smell-specific characteristics \emph{Shape} and \emph{Affected design level} are of interest because they are directly related to the complexity of the smell.
The more complex the shape, and the more edges there are between the affected components, the harder the smell is to refactor because more effort is required.
Similarly, the affected design level is important because the cycles present at both component/package and file/class level have an impact on two different abstraction levels at once.

The UD smell-specific characteristics \emph{Instability gap} and \emph{Strength} are of interest because they are used for the detection of the smell and thus can effectively measure its criticality. The higher the \emph{Instability gap}, the higher the chance the component affected by the smell is changed due to ripple effects \cite{martin1994}. Likewise, the higher the strength, the higher the chance (because there are more possible components that are prone to a change) that a change will occur and propagate to the affected component.

The HL smell-specific characteristic \emph{Affected ratio} is of interest because it quantifies the involvement of the files that belong to the affected component.
The higher the affected ratio, the stronger the connection between afferent and efferent components and the higher the coupling.
The \emph{Afferent ratio} and \emph{Efferent ratio} divide this concept by only counting the files with incoming or outgoing dependencies from/to external components in the central component, respectively.
Thus, they help to understand whether the central component provides more or less services than it uses itself. These two characteristics are basically a breakdown of the affected ratio.

The only GC smell-specific characteristic that we study in this work is the density of the lines of code.
The denser the GC, the more code the files in the component have, meaning that those files are harder to maintain due to a larger size in terms of lines of code \cite{Nunez-Varela2017}.

\begin{table}[tbp]
    \footnotesize
    \centering
    \caption{The architectural smell characteristics analysed in this study.}\label{tab:characteristics}
    \begin{tabular}{p{0.25\linewidth}|p{0.69\linewidth}}\toprule
        \textbf{Name} & \textbf{Description} \\ \midrule
        \multicolumn{2}{c}{\itshape smell-generic} \\ \midrule
        Age & The number of versions the smell is present in. \\
        Size & The number of artefacts affected by the smell. \\
        Centrality & The importance of the artefacts affected by the smell within the dependency network of the system.  Measured using PageRank (from on \cite{Roveda2018b}). \\
        Number of edges & The number of dependency edges among the affected artefacts. \\ \midrule
        \multicolumn{2}{c}{\itshape CD smell-specific}\\\midrule
        Shape & The shape of the cycle: tiny, circle, chain, star, clique (from \cite{AlMutawa2014}). \\
        Affected design level & Whether the cycle is present only among files or components or at both levels (from \cite{AlMutawa2014}). \\ \midrule
        \multicolumn{2}{c}{\itshape UD smell-specific}\\\midrule
        Strength & The ratio between the number of dependencies that point to less stable components and the total number of dependencies of the class (from \cite{Fontana2016}). \\
        Instability gap & Is the difference between the instability of the main component and the average instability of the dependencies less stable than the component itself  (from \cite{Fontana2016}). \\\midrule
        \multicolumn{2}{c}{\itshape HL smell-specific}\\\midrule
        Affected ratio & The ratio between the number of files creating the central component's incoming and outgoing dependencies and the total number of files in the central component (based on \cite{Abdeen2011}). \\  
        Afferent ratio & The ratio between the number of files within the central component with incoming dependencies from external components and the total number of files within the central component (based on \cite{Abdeen2011}). \\ 
        Efferent ratio & The ratio between the number files within the central component with outgoing dependencies to external components and the total number of components the central component (based on \cite{Abdeen2011}). \\\midrule
        \multicolumn{2}{c}{\itshape GC smell-specific}\\\midrule
        LOC Density & The total number of lines of code present in this component divided by the number of files in the component (i.e. its Size). \\ \bottomrule
    \end{tabular}
\end{table}

\subsection{The \textsc{Arcan} tool}
To detect AS, we extended \textsc{Arcan} to support the proprietary C/C++ used by the company participating in this study. 
\textsc{Arcan}'s results were validated by previous studies and obtained a precision ranging from 70\% to 100\% \cite{Fontana2020, Fontana2017}.

\textsc{Arcan} parses Java, C, and C++ source code files to create a dependency graph where files, components, classes and packages are all represented using different nodes with different labels. Dependencies, and other relationships between nodes, are represented using edges that connect the dependant to its dependencies with an outgoing, labelled edge (e.g. if artefact \texttt{A} depends on artefact \texttt{B}, then the dependency graph contains a directed edge connecting \texttt{A} to \texttt{B}.). 
The project's structural information contained in the dependency graph is then used to calculate several software metrics (e.g. fan-in, fan-out, instability \cite{martin1994}, etc.) and then detect architectural smells by recognising their structure in the dependency graph.

Compared to other tools, \textsc{Arcan} uses only software metrics and structural dependencies in order to detect architectural smells. 
This makes \textsc{Arcan} different from tools such as DV8 \cite{Mo2015} (a tool used by related work) which also requires the use of change metrics.
While this type of metrics definitely provide important insights into the maintenance hotspots of the system, they also come with the requirement of needing historical data in order to be used.
This aspect is of particular importance in our case as the version control system adopted by the company we worked with, \emph{did not provide} such information.

Despite the different approaches to detect architectural smells, the two tools,  \textsc{Arcan} and DV8, have some overlap in the detected smells. 
Both tools detect cycles among files and components and both detect hub structures (called \emph{Crossing} by DV8 and Hublike Dependency by \textsc{Arcan}), but DV8 also incorporates historical information for the detection of the latter type of smell.

\subsection{Similarities and differences between code and architectural smells}
Distinguishing between code smells (CS) and AS may not always be easy as different authors have a different understanding of what constitutes one or the other.
In this section, we try to provide a brief explanation about both and clarify the differences between these two concepts.

Code smell is a term first popularised by Kent Beck in the late 1990s \footnote{Read \url{https://wiki.c2.com/?CodeSmell} for more info.} and then further defined by Martin Fowler and Kent Beck himself in the early 2000s \cite{Fowler2002}.
A CS is a sign that the piece of code under inspection requires some changes (i.e. a refactoring) in order to be considered of good quality \cite{Fowler2002}. 
In other words, code smells are symptoms of poor design and implementation choices \cite{Tufano2015}.

The term architectural amell was first adopted by Lippert \cite{Lippert2006} in 2006 to describe a part of the system that required significant refactorings at the architecture level in order to meet the desired quality standards.
To be more specific, Lippert mentions that architectural smells, contrary to code smells, require \emph{large refactorings} in order to be removed from the part of the system they affect and require longer than a day to be applied \cite{Lippert2006}.

Both AS and CS manifest themselves in different forms that are commonly referred to as different \emph{types}.
Some examples of CS types are \emph{Duplicated Code}, \emph{Long Method}, and \emph{Large Class} \cite{Fowler2002}.

Finally, it is important to mention that previous work provides empirical evidence that the AS  considered in this study and the most well-known code smells are \emph{independent} entities and that there is no correlation between the presence of AS and CS \cite{Fontana2019}.

\section{Related Work}\label{sec:related-work}
This section summarises similar studies from the literature regarding architecture smells and (to a lesser extent) code smells.

In our previous study \cite{Sas2019}, we investigated the evolution of AS in open source Java systems by adopting two techniques from other domains (that were previously applied in software engineering): Dynamic Time Warping and Survival Analysis. Specifically, we examined how a set of AS characteristics evolve and how long AS survive within the system.
Our findings showed that Cyclic dependencies have a low survival rate (just a few weeks for more than 50\% of instances), and Hublike Dependencies are much more complex than cycles. In general, this means that Hublike Dependencies are a much better option for refactoring than cycles. 
The present study is different from our previous work because it focuses on industrial C/C++ embedded systems and it investigates the opinions of the architects and developers working on the analysed projects.

Martini et al. \cite{Martini2018} studied the relationship between AS and Architectural Technical Debt (ATD) within an industrial partner. 
They used questionnaires and focus groups to collect the opinion of practitioners concerning a selected set of architectural smells detected in four Java projects.
Their findings showed that practitioners were not aware of half of the smells detected in their systems. Furthermore, those practitioners ranked AS in terms of their cost to refactor, placing Cycles first, followed by Hublike Dependencies, and then by Unstable Dependencies.
Our study differs from Martini et al.'s study because we analyse C/C++ projects from an embedded systems company and use individual interviews to collect qualitative data.
Additionally, we focus on analysing the evolution of architectural smell instances, collect the experiences of architects and developers dealing with those smells, and their opinion on the results.
Martini et al., instead, perform a qualitative analysis aimed at prioritising the refactoring of the smells detected and try to understand architectural smells' impact on ATD.

Arcelli et al. \cite{Fontana2020} performed a similar study to Martini et al. but in a different industrial setting and extended the study to 8 different types of smells (instead of only 3).
Their findings highlight that practitioners recognise the impact of AS on Maintainability, but were not aware of the definition of many of the 8 types of smells investigated.
Similarly to Martini et al., practitioners recognised Hublike Dependency as a primary candidate for refactoring, and mentioned that some smell types (Feature Concentration, Scattered Functionality and Insufficient Package Cohesion) are only useful to consider in a layered architecture.
This work differs from our study because we analyse C/C++ projects and used individual interviews rather than a survey to collect the developers' opinion. Moreover, we also combine quantitative and qualitative data instead of focusing only on the latter. 
For example, Arcelli et al. focus on how architectural smells refactoring is approached by practitioners, and while we partially cover this topic too in our interviews, we also show our subjects smell instances detected in their system on which they can base their answer on.
Finally, our study puts a strong emphasis on the maintenance and evolution issues related to architectural smells as experienced by practitioners, while Arcelli et al.'s focuses on how practitioners perceive architectural smells in general. 

Mo et al. \cite{Mo2018} performed an industrial study to measure the maintainability of the architecture using two metrics and the architectural ``hotspots'' that incur high maintenance costs within 8 C/C++ and C\# projects from a large software company.
The authors also complemented their analyses with interviews with 6 subjects working for the company they collaborated with.
Their findings confirm that the tool suite they used is instrumental for architects to pinpoint, visualise, and quantify ``hotspots'' in the architecture of the system.
Similar to results in other studies \cite{Fontana2020}, the development teams mentioned that they were mostly aware of the key problems affecting their system, but it was usually hard for them to specify or quantify those problems.
In terms of the research method, the study of Mo et al. is similar to ours, as both studies feature a collaboration with a large software company where a tool was used to create a report and present it to practitioners in order to collect their feedback.
Our work differs from the one of Mo et al. in two key aspects: (1) our study is more specific and focuses on a different set of architectural smells while Mo et al. combine three different types of analyses, two of which do not concern architectural smells; and (2) we focus specifically in studying the evolution of architectural smells in industrial systems, while Mo et al. focus on the overall experience of applying an automated tool suite in an industrial context.

De Andrade et al. \cite{DeAndrade2014} investigate the architectural smells defined by Garcia et al. \cite{Garcia2009} in an open source software product line (SPL) written in Java.
Their study is mostly exploratory in nature and focuses on how architectural smells affect SPLs by performing a manual detection of architectural smells using a reverse-engineered component model of the SPL.
Their findings mostly provide insights about the SPL under analysis and the specific instances affecting it.
Our study differs from De Andrade et al. because we look at the evolution of smell instances over time rather that at the implications created by architectural smells at a single point in time.

Nayebi et al. \cite{Nayebi2019} performed a longitudinal study on how the architectural smells detected in an industrial Java system changed after a comprehensive refactoring of the system. 
The authors analysed the system in question 6 months before and 6 months after the refactoring took place.
Their findings show that the average time needed to close issues was reduced by 72\% as well as the number of lines of code needed to do so.
The authors also performed two interviews to collect qualitative data from two key actors of the company.
Their findings show that the reports describing the amount of architecture debt present in the system were crucial to convey to the top management the necessity of performing refactoring.
Our work differs in both its scope and goals.
The scope of our paper was a large multinational company that mostly adopts C/C++, whereas Nayebi et al. collaborated with a start-up company that operates worldwide and works with Java.
The goal of our study is to understand how individual instances of smells evolve over time in industrial systems and how their effects are perceived by architects and developers, whereas Nayebi et al. aimed at studying the effects of refactoring on architectural technical debt (using architectural smells as proxy).

Feng et al. \cite{Feng2019} studied how three change propagation patterns, identified by the authors, affect the components involved in the architectural smells detected by the DV8 tool. 
Their findings show that there exist only a few dominating active hotspots in the evolution timelines of the 21 Java OSS projects they considered.
Our study differs from their work because we focus on the evolution of the individual instances rather than on the change patterns generated by these.
Moreover, we also collect qualitative data concerning the perceived effect of architectural smells by C/C++ industrial practitioners.

Xiao et al. \cite{Xiao2016} studied how an architectural technical debt index can be modeled using architectural smells and statistical models. Their findings show that the top 5 architectural smells (or architectural debts, using the terminology of the authors) consume a large amount of the total project effort spent on maintenance.
Our study differs from their work in terms of focus and scope. 
The focus of our study is understanding how architectural smell instances evolve in the scope of industrial C/C++ projects, whereas Xiao et al. focused on the relation between architectural smells and historical changes to the files affected; they also modelled this relation and summarised it as an index.
Moreover, they also focus on a different set of architectural smells and work with open source projects.

Other similar studies from the literature focus on CS, rather than on AS. However, CS are different entities than AS, as empirically verified in a previous study on the matter \cite{Fontana2019}. Thus, we only briefly summarise two of them here because of the similarity in the data analysis methodologies.

Palomba et al. \cite{Palomba2018} investigated the co-occurrence and introduction order of code smells in open source Java systems, finding that more than 50\% of smelly classes are affected by more than one smell and that method-level smells may not be the root cause of the introduction of class-level smells.
We used similar techniques to Palomba et al. to analyse the introduction order and co-occurrence of architectural smells.

Finally, Vaucher et al. \cite{Vaucher2009} tracked a design smell (God Class) in order to understand whether the smell originated with the class (i.e. it is by design), or occurred by accident (i.e. it is considered bad code).
The findings show that the God Classes that are by design are less likely to be changed from version to version, contrary to classes that become God Classes over time.
Our approach to classify the trend of smell characteristics over time is inspired from the approach of Vaucher et al. to track God Classes.


\section{Case study design}\label{sec:case-design}

\subsection{Goal and Research Questions}
The research goal of this study is to improve the current knowledge on architectural smells evolution within a system and understand how practitioners perceive their presence in terms of consequences on Maintainability and Evolution.
Using the Goal-Question-Metric \cite{VanSolingen2002} approach, the goal can be formulated as: 
\begin{quote}
    \itshape
    \textbf{Analyse} architectural smell instances throughout a system's history \textbf{for the purpose of} understanding how they evolve and are perceived by practitioners \textbf{with respect to} their characteristics, lifespan, co-occurrence, and introduction order \textbf{from the point of view of} software architects and engineers \textbf{in the context of} industrial software systems.
\end{quote}

The goal is further refined into five research questions. For each research question we explain its purpose and how it helps to advance the state of the art.

\begin{description}
    \item[\textbf{RQ1}] \textit{How do architectural smells evolve in industrial software systems?}
     \begin{description}
         \item[\textbf{RQ1.1}] \textit{How do their characteristics evolve over time?}
         \item[\textbf{RQ1.2}] \textit{How long do different smell types persist within the system?}
     \end{description} 
\end{description}

This question is answered by answering the two sub-research questions.
The first sub-research question aims at investigating the changes that occur in the individual instances of architectural smells in terms of the smell characteristics (e.g. their size, their centrality, etc. -- see Section \ref{sec:characteristics}). This will allow us to understand what aspects of a smell change over time, and more generally, how the changing structure of a smell affects the system over time.
The second sub-research question focuses on understanding the survival rate of different smell types within the system as it evolves. This will allow us to understand in depth what smell types influence Maintainability the most on the long-term by simply having more time to influence the system. Subsequently, this can help to define new or refine existing prioritisation techniques for architectural smell refactoring.

\begin{description} 
    \item[\textbf{RQ2}] \textit{What pairs of architectural smell types co-occur more often?}
\end{description}
This question investigates the co-occurrence of different smell types in the same software component (e.g. class, or package; file, or folder). The answer to this research question can provide insights on what pairs of smells tend to appear together often. Such insights can subsequently help in reducing the number of smells introduced by alerting developers in advance of the possibility of performing some preemptive refactoring.

\begin{description}
    \item[\textbf{RQ3}] \textit{What architectural smell types are more likely to precede or succeed other smells in co-occurrences?}
\end{description}
This research question is a follow-up to the previous one. It focuses on uncovering what smell types temporally precede or succeed other smell types.
Such information can be used to notify developers that the presence of a certain instance is likely to lead to the introduction of more smells of a different type, therefore allowing them to take appropriate measures. Spending some effort to remove an architectural smell, can yield a great return of investment, if it prevents extra maintenance and rework due to multiple other smells appearing in the future.

\begin{description}
    \item[\textbf{RQ4}] \textit{How does information about architectural smell evolution help practitioners?} 
\end{description}
The goal of this research question is to find out if and how information about AS helps practitioners in identifying and understanding problems in their architecture, whether they are aware of these problems in the first place, and what aspects of the analysis are the most helpful (e.g. historical data, smell characteristics, summary of the analysis, etc.).
We ask this RQ to examine how useful the information on architectural smell evolution is in practice for reducing maintenance effort. 
This also entails understanding what parts of the analysis provide the most interesting and actionable insights to practitioners.
Additionally, this RQ might uncover if there is any key information missing from what is reported to the developers.

\begin{description}
    \item[\textbf{RQ5}] \textit{How do architectural smells impact a system's Maintainability and Evolvability?}
\end{description}
This research question investigates the effects of AS on maintenance and evolution as perceived by software practitioners.
More specifically, the RQ studies the aspects that decrease the Maintainability level of the affected parts, the long-term development of new features (i.e. Evolvability), the possible quality-improvement strategies practitioners might consider, and what information would help them implement those strategies best.
Ultimately, this information can be of great importance in improving the quality of the output offered by tools that automatically detect and analyse AS. The difference between RQ4 and RQ5, is that the former deals with the problems in the architecture (smells per sé), while the latter deals with the consequences of those problems (on maintenance and evolution) as well as how to solve them.

To facilitate reproducibility, we provide a replication package for this study\footnote{Visit \url{https://doi.org/10.6084/m9.figshare.16884739.v1}.} 
containing the study protocol, the R scripts used for data analysis, and many other resources.

\subsection{The necessity of studying AS in an industrial setting}
To the best of our knowledge, the vast majority of studies on this topic have a \emph{limited scope} and only focus on open source systems that are mostly written in Java, or focus on a different set of AS.
This limits our understanding of how architectural smells actually impact the work of practitioners in real world scenarios.
Moreover, this only allows a narrow perspective based on quantitative results thus overlooking the (usually more nuanced) qualitative data. 
More specifically, it is of interest to understand how developers and architects are affected by the presence of architectural smells, whether they are aware of the problems in the first place, and if so, what decisions they make in order to remedy such problems.

Furthermore, Java systems are characterised by several different types of dependencies (e.g. call, inheritance, use, etc. \cite{Pruijt2017}) and provide constructs such as polymorphism that offer programmers several ways to interconnect classes and interfaces and create dependencies among them.
Procedural languages such as C, on the other hand, have a limited set of built-in features and do not encourage the creation of dependencies as much as their OO counterparts.
Moreover, as we will explain over the next sections, the company we collaborate with has developed proprietary mechanisms for defining dependencies between components which might alter the way we interpret dependencies and thus all architectural smells, the detection of which is based on dependencies (CD, UD, and HL).

\subsection{Research Method}
To achieve the aforementioned goal and answer the five stated research questions, we collaborated with a large technology industrial partner, ASML\footnote{Visit \url{www.asml.com} for more info.}, to analyse a few of their projects and interview some of the engineers working on these projects. 

\begin{figure}[H]
    \centering
    \includegraphics[width=.7\textwidth]{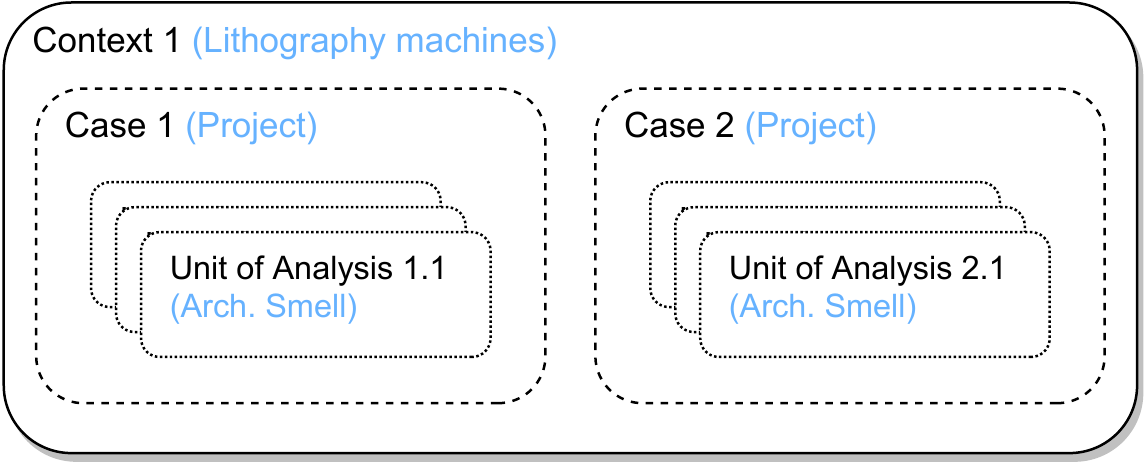}
    \caption{The case study design using Runeson et al.'s representation \cite{Runeson2012}.}
    \label{fig:case-design}
\end{figure}

More precisely, the company showed interest in analysing one massive software product line (of 20 million LOC) that is composed of multiple projects.
The projects are primarily written in C/C++ and compiled using a proprietary compiler and auxiliary tools.
The main business of the company is manufacturing industrial machinery for the mass production of microchips. Therefore, all the projects considered in this study belong to this domain. In terms of our case study, the projects are designated as the cases, and the units of analysis are the architectural smells detected in each project. Figure \ref{fig:case-design} illustrates the case study design.

The five stated research questions require different types of data in order to be answered.
RQ1, RQ2, and RQ3 necessitate \emph{quantitative} data about architectural smells extracted from multiple versions of the source code of each project.
RQ4 and RQ5, instead, require \emph{qualitative} data collected from software architects and engineers working on the studied projects.
The remainder of this section explains how the data collection for these two groups of research questions was performed.

\subsubsection{Quantitative data collection}
\paragraph{Projects and Architecture}
The first step of performing \emph{quantitative} analysis is selecting the cases to analyse. 
The selection of the projects was done in consultation with an architect from the company. We requested that the list of projects would differ as much as possible in terms of total number of lines of code (LOC), to maximise the diversity in our sample.
The selection was also influenced based on the interest of the architects responsible for each project in obtaining information regarding the presence of architectural smells in their systems.
The final list of projects is shown in Table \ref{tab:projects}.
The projects differ greatly in total number of lines of code analysed, from a few thousands to a few million. Each project is also responsible for a single step in the manufacturing process of the microchip.
One project (P09) is relatively new compared to the rest, and thus smaller both in terms of LOC and number of versions.
We also note that, over time, the company has split projects in two or more parts to better manage them, causing both a steep decrease in the LOC of some projects, and other projects starting with a high number of lines of code.

It is important to mention a few details about the architectural style of the projects selected. 
The company adopts a layered architectural style with each project (ideally) only communicating with projects from layers below them or from the same layer.
Each project is divided into multiple clusters of components that handle a specific functionality provided by that project. Larger projects may be divided into multiple teams, each maintaining their own cluster of components.
Projects situated in higher layers provide functionality that allow the user to command the machine and configure it.
In contrast, projects located in lower layers are responsible to govern the hardware, orchestrate other components, and provide abstraction layers to allow the deployment of the code on different kinds of hardware.
Finally, we note that all projects contain both C and C++ files, with the former type being the most popular one. 

\begin{table}[tbp]
    \footnotesize
    \centering
    \caption{The list of projects analysed in this study. MLOC = Million Lines Of Code}
    \label{tab:projects}
    \begin{tabular}{@{}llll@{}}
    \toprule
    \textbf{ID} & \textbf{Description} & \textbf{Versions analysed} & \textbf{MLOC last version} \\ \midrule
    P01  & Reticle handling     & 31        & 4.88         \\ 
    P02  & Waterflow control    & 37        & 3.92         \\ 
    P03  & Dose control         & 37        & 1.58         \\ 
    P04  & Light control        & 37        & 2.22         \\ 
    P05  & Immersion control    & 37        & 0.86         \\ 
    P06  & Device \& Data subsystems & 37    & 6.45         \\ 
    P07  & Machining control    & 37        & 2.56         \\ 
    P08  & Input data manager   & 18        & 2.31         \\ 
    P09  & Alignment \& Diagnostics & 9     & 0.011      \\ 
    \bottomrule
    \end{tabular}
\end{table}

\paragraph{Architectural Smells Detection}
The analysis of the projects included the following phases: detection of architectural smells, the tracking of architectural smells over time, and the calculation of the software metrics necessary for the data analysis. 
The detection-tracking process is repeated for every version available and the results are merged at the end of the whole process.

The detection of smells is carried out using \textsc{Arcan}, which generates a dependency graph (DG) given the source files of a C/C++ or Java project.
The DGs of these two languages, however, present several differences that influence the architectural smell detection process.
For example, DGs for C/C++ projects have nodes and edges that respectively represent and connect header files, which are obviously not present in DGs of Java projects.
Moreover, the package structure of Java projects is a tree structure that requires dependencies to propagate from the leaves (i.e. the classes) to their parents (i.e. the packages containing those classes, the packages containing those packages, and so on).
In ASML, however, there is no such concept and there are no child components.
The different structures of these two languages (or, more technically, the different graph schemas) imply that dependencies are constructed differently: in the case of this study, the detection of architectural smells was tailored based on the guidance of ASML engineers.
In particular, components were treated as packages and header files as Java interfaces, but only for the purpose of mining dependencies (i.e. headers were not considered for smell detection).
All the dependencies detected in the header files were carried over to the exact files implementing, or using, those dependencies.

Note that, in order to extract the dependency graph, we had to write specific code that would account for all the proprietary changes the company implemented to their compiler, and consequently to the syntax of the code. 
Additionally, since some of the files were automatically generated at compile time, we were also required to compile the projects in order to pick up as many dependencies as possible. These files contained dependencies between internal components that were manually declared by the engineers in a proprietary file format, and missing these dependencies would have eventually resulted in incomplete results.
These two tasks turned out to be very time-consuming, and packed with arduous technical challenges.

The tracking of the smells is then done using \textsc{ASTracker} \cite{Sas2019}, which matches smell instances from two adjacent versions that correspond to the same smell (i.e. they affect the same files but in adjacent versions).
Usually, this process is susceptible to file renamings; however, the file naming policies of the company prevented the introduction of noise in this part of the analysis, as file renamings are not an encouraged practice.

The versions we analysed were all the snapshots of the projects that the company tagged as \emph{releases} in their version control system (VCS). The time period taken into consideration is 3-years long (from 2017 to 2020) and each release took place, on average, 35 days after the previous. Note that the we stopped at 3 years because the VCS used by the company at the moment was adopted 3 years before the start of this research.

\begin{figure}
    \centering
    \includegraphics[width=1\linewidth]{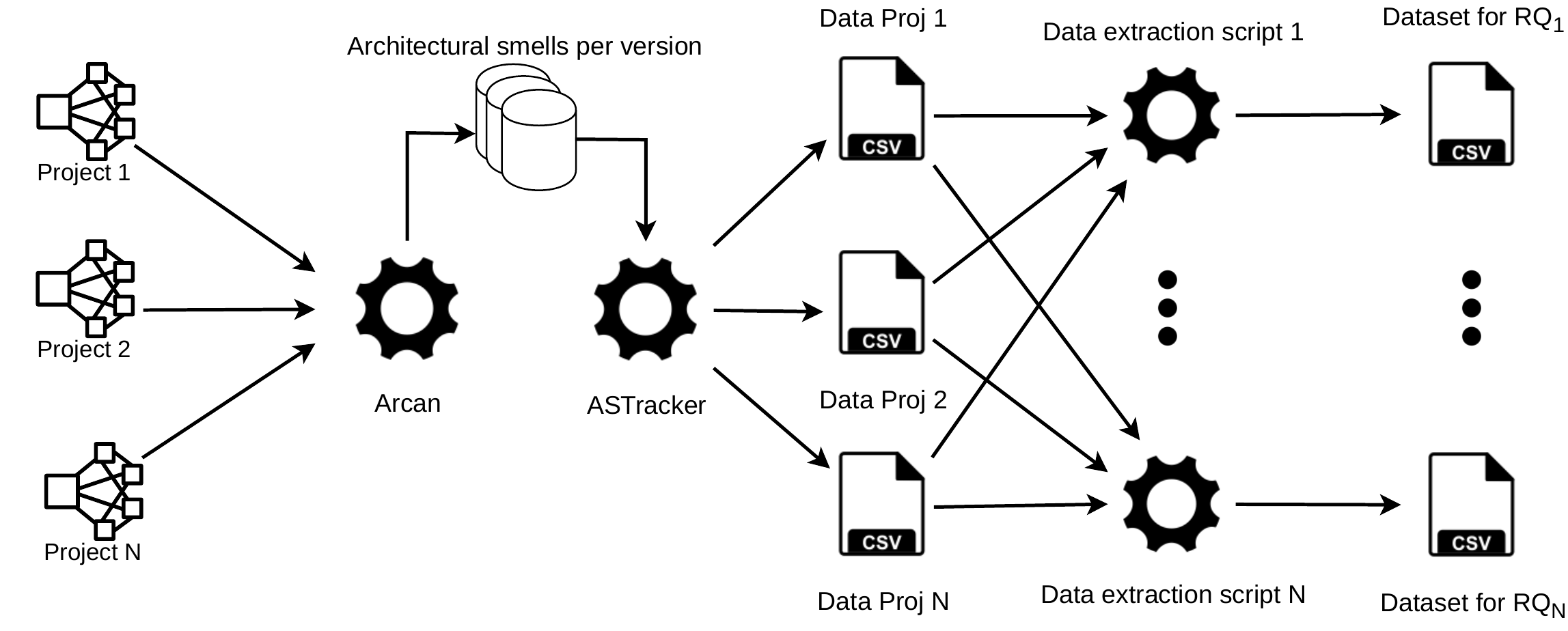}
    \caption{Quantitative data collection process.}\label{fig:data-collection-process}
\end{figure}

A detailed representation of the whole data collection process is shown in Figure \ref{fig:data-collection-process}.
For each version in the VCS, we compiled the source code to obtain the automatically generated files (omitted from Figure \ref{fig:data-collection-process}), then we ran \textsc{Arcan} on each project to obtain the dependency graph of that version.
At the end of the analysis, we ran \textsc{ASTracker} to synthesise the information contained in the dependency graphs into CSV files of raw data. These files were then processed to create the datasets for each individual research question.

\subsubsection{Qualitative data collection}
While RQ1, RQ2 and RQ3 required the \emph{quantitative} data described in the previous sub-section, RQ4 and RQ5 required \emph{qualitative} data to be fully answered. 
To this end, we planned a series of \textbf{interviews} with the engineers and architects working on the projects we analysed.

The process for selecting the participants to our interviews started with a presentation of our analysis in one of the monthly meetings between all the architects of the company.
Architects that showed interest were contacted and their projects were analysed. 
Afterwards, we prepared an interactive report\footnote{An anonymised version is available in the replication package of this study.} specific to each project analysed and sent it to the corresponding architect.
Each architect was then asked to pick a handful (3-5) of engineers that we could interview; they were also asked whether they would like to take part in the interview themselves.
Each participant received a consent information letter, informing them of their rights as participants, and a copy of the report with the results of the analysis. The report also contained a quick guide to allow them to understand the results. The participants were asked to inspect the report before taking part in the interview. 

\begin{figure}
    \centering
    \includegraphics[width=1\linewidth]{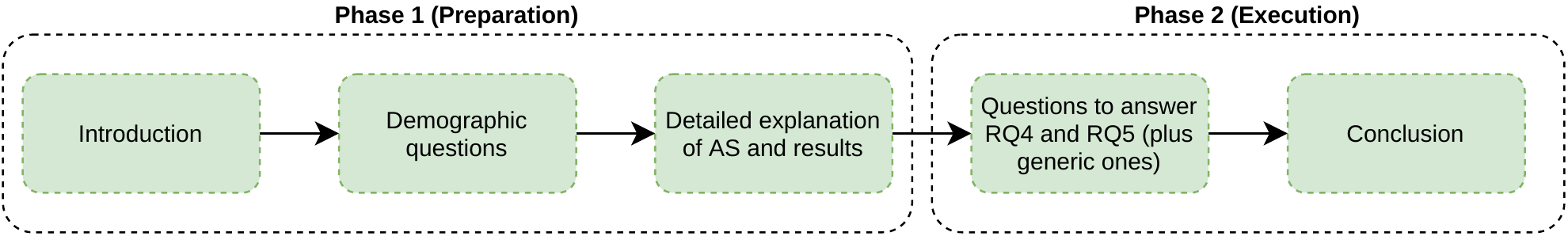}
    \caption{The phases and structure of the interviews.}\label{fig:interview-phases}
\end{figure}

\begin{figure}
    \centering
    \includegraphics[width=1\linewidth]{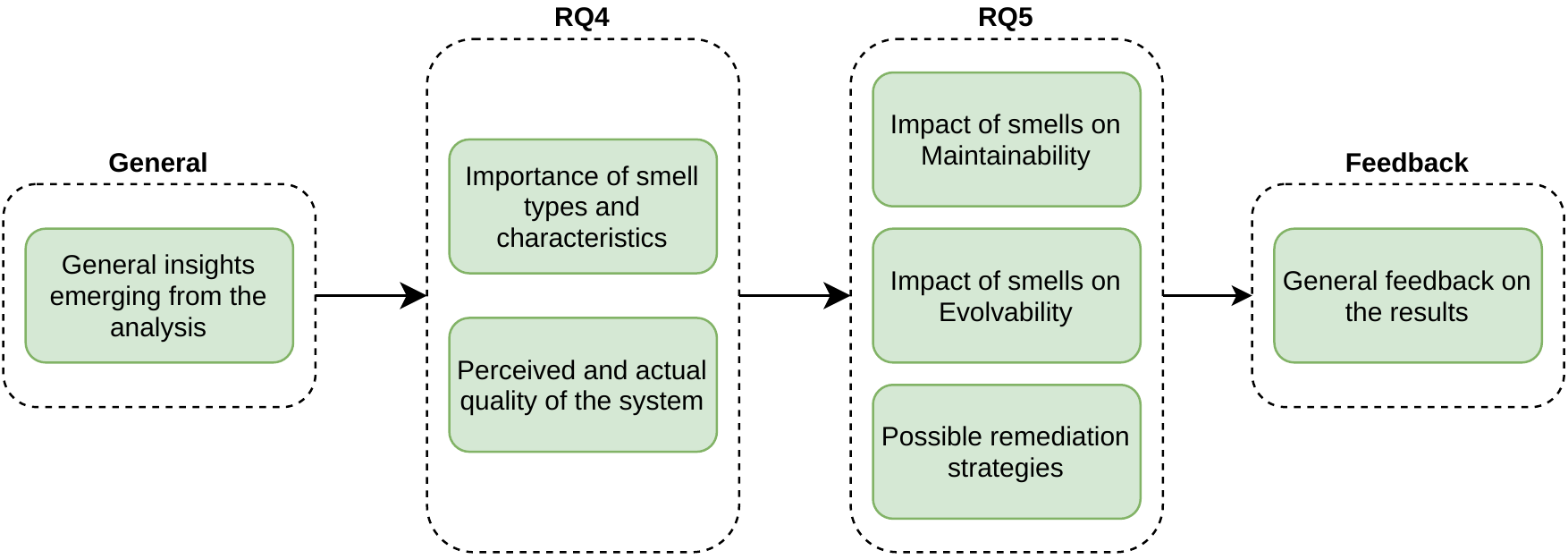}
    \caption{The structure of the first step of phase two, focusing on RQ4 and RQ5.}\label{fig:interview-questions-sub-phases}
\end{figure}

The interviews lasted 30 to 40 minutes each and were performed remotely by the first author using video-conferencing, individually with each participant listed in Table \ref{tab:participants}.
Interviews followed a semi-structured format \cite{Runeson2012} as depicted in Figure \ref{fig:interview-phases} and further detailed in Figure \ref{fig:interview-questions-sub-phases}.
As it can be noted, the actual questioning session (Phase 2, in Figure \ref{fig:interview-phases}) was preceded by an introduction to the study, some demographic questions, and an explanation of the key theoretical concepts necessary to understand the questions in Phase 2.
The questions asked in Phase 2, were grouped by topic and map to either RQ4 or RQ5, as shown in Figure \ref{fig:interview-questions-sub-phases}.
Given the semi-structured format, the interviewers also asked follow-up questions and may have not followed the predefined list of questions if an interesting point, worth of further investigation, was touched during the session.
The full interview guide is available in Appendix \ref{appendix:interview-guide}.

\begin{table}[tbp]
    \centering
    \caption{Background information on the interviewee and their respective projects.}
    \label{tab:participants}
    \begin{tabular}{@{}clm{4cm}cc@{}}
        \toprule
        \multirow{2}{*}{\bfseries ID} & \multirow{2}{*}{\bfseries Project} &\multirow{2}{*}{\shortstack{\bfseries Official position}} & \multicolumn{2}{c}{\bfseries Years of exp.} \\
        &  &  &\multicolumn{1}{c|}{\bfseries curr. role} & \textbf{in total} \\ \midrule
        I0 & P05 & Product Architect & 4 & 8 \\    
        I1 & P03 & Design Engineer & 3 & 4 \\    
        I2 & P03 & Software Architect & 8 & 14 \\   
        I3 & P03 & Design Engineer & 4 & 15 \\   
        I4 & P03 & Design Engineer & 2 & 8 \\    
        I5 & P08 & Design Engineer & 4 & 10 \\   
        I6 & P08 & Software Architect & 6 & 15 \\   
        I7 & P07 & Software Architect & 7 & 25 \\   
        I8 & P07 & Software Architect & 8 & 22 \\   
        I9 & P07 & Software Architect & 6 & 23 \\   
        I10 & P02 & Design Engineer & 3 & 3 \\   
        I11 & P02 & Lead Design Engineer & 5 & 20 \\  
        \midrule
    \multicolumn{3}{l}{\textit{\textbf{Average}}} & 5 & 13.9 \\ \bottomrule
    \end{tabular}
\end{table}

\section{RQ1 -- Architectural smells evolution}\label{sec:rq-1}
\subsection{RQ1.1 -- Evolution of smell characteristics}\label{sec:rq1.1}
\subsubsection{Data analysis methodology: Dynamic Time Warping}\label{sec:methodology-rq1.1}
To understand how smell characteristics evolve over time, we adopt the same technique we used in our previous work \cite{Sas2019, Vaucher2009}: signal classification with Dynamic Time Warping\footnote{The implementation used for this analysis was provided by the R package \texttt{dtw}.} (DTW) \cite{Kruskal1983}.
This approach considers every series of values of every characteristic of every smell instance as a signal (or time series) and then compares each signal to a series of predefined signals (templates), each one with a corresponding label. 
Depending on the template that is the mathematically closest to the signal, a label is assigned to it.

\begin{figure}[H]
    \centering
    \includegraphics[width=0.9\textwidth,trim={0.2cm 0 0 0},clip]{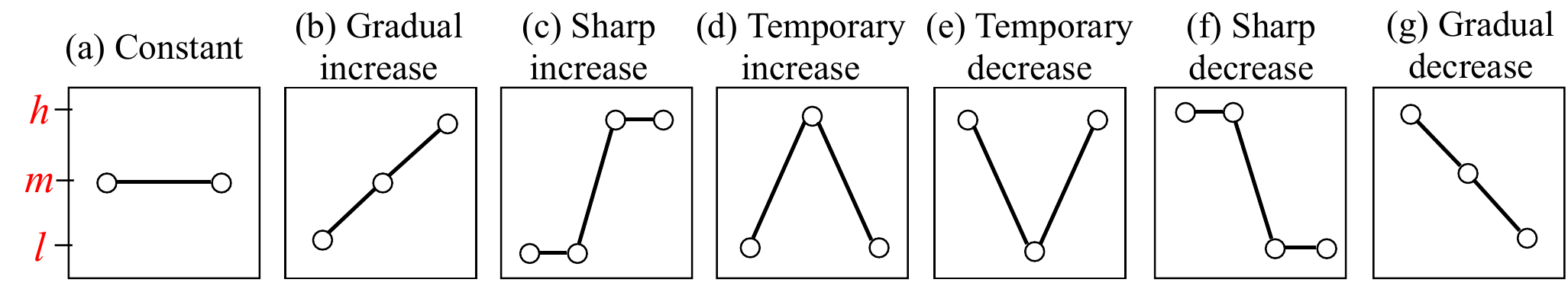}
    \caption{Trend evolution classification templates. Figure adapted from the work of Vaucher et al. \cite{Vaucher2009}.}\label{fig:classification-templates}
\end{figure}

Formally, we can model the problem as follows: for every smell characteristic $C^{k}$ of a certain smell $k$ we consider the different values $C^{k}_i$ as a signal $S$. We then compute the following variables: $h = \max S$; $l = \min S$; and $m = (h+l)/2$.
These three values are then used to build the seven templates, named from $a$ to $g$, shown in Figure \ref{fig:classification-templates}. For example, template (c) is defined as $c = (l, l, h, h)$.
The values $l$, $m$, $h$ are re-calculated for each signal classified.
Finally, the signal is classified by comparing the distance of the signal from each template, and selecting as a label the name of the closest template. 

Even though the selected templates offer a good variety of possible signal shapes, there exist some cases that may not be well approximated by the current selection.
One example is a signal that varies between two integer values (e.g. 6-7) multiple times, which would be classified by the model as a constant signal (i.e. template (a)).
Nonetheless, we deem that the approximation offered by the model when classifying such unusual signals, is sufficient for the purpose of this paper for the following reasons:
\begin{itemize}
    \item the templates selected represent simple and general cases, thus they simplify interpretation and analysis;
    \item a signal is classified based on the distance DTW calculates between the points from the template and points from the signal, thus the classified signal has at least an internal component that resembles the assigned template.
\end{itemize}

\subsubsection{Results}\label{sec:results-rq1.1}
The results shown in this section concern smells that affected the system for at least 3 releases, in order to avoid spurious outcomes and focus on long-lived smells. Note that in this section, unless specified, or the context implies otherwise, when we refer to an AS instance we usually mean a smell that was detected in multiple versions and was identified as the same smell.

Finally, we use the following terminology: \emph{version} and \emph{release} are used interchangeably, \emph{component} refers to a group of files defined as such by the architects of the system, \emph{artefact} refers to both files and components, and the terms \emph{co-occurrence} and \emph{overlap} (among AS) are used interchangeably.

\begin{table}[tbp]
    \centering
    \caption{The number of unique temporal AS instances that that have an age of at least 3.}
    \label{tab:smell-count}
    \begin{tabular}{@{}lcc|c@{}}
    \toprule
    \textbf{Smell Type} & \textbf{File-level} & \textbf{Component-level} & \textbf{Total} \\ \midrule
    Cyclic Dependencies & 14637 & 941 & \textbf{15578} \\
    Hublike Dependencies & 151 & 40 & \textbf{191} \\
    Unstable Dependencies & -- & 273 & \textbf{273} \\
    God Component & -- & 190 & \textbf{190} \\ \bottomrule
    \end{tabular}
\end{table}

\begin{figure}[h]
    \centering
    \includegraphics[width=\textwidth]{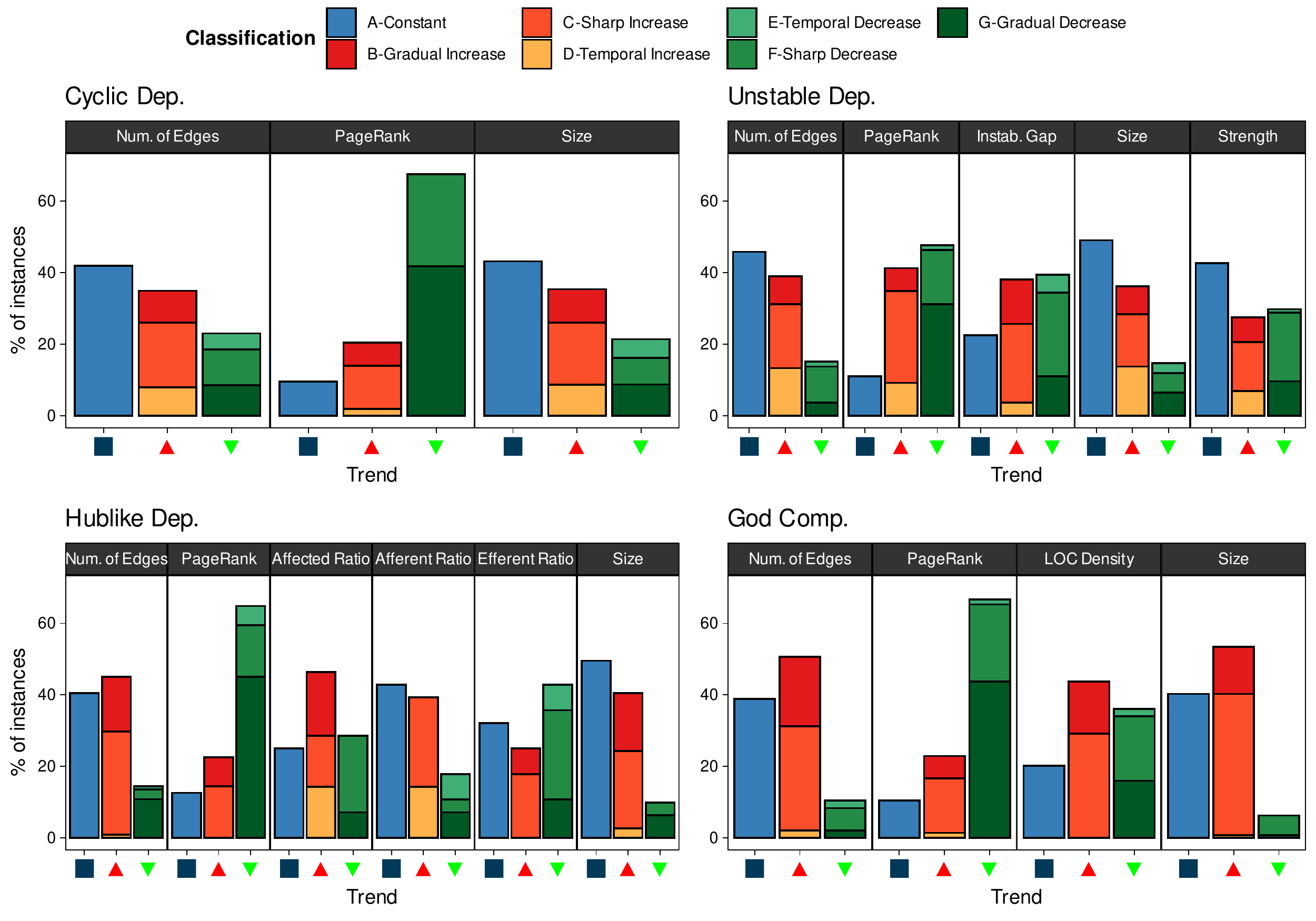}
    \caption{The classification of AS's instances trend for various characteristics grouped by smell type. The classification is in percentage of the total number of instances with at least an age of 3. The classification (represented by the colour) for each characteristic is grouped by the overall trend (constant, represented as a black block; increasing, represented as an up-pointing triangle; and decreasing, represented by a down-pointing triangle).}\label{fig:dtw-classification}
\end{figure}

\paragraph{Cyclic Dependencies (CD)} As Figure \ref{fig:dtw-classification} shows, most of the 15578 CD instances exhibit either an increase in the number of artefacts affected (i.e. size) or they remain steady over time. More specifically, 43\% increase in size in some way, 36\% stay constant, and only 21\% of them decrease. As expected, a similar pattern also emerges when looking at the number of edges among the affected artefacts (since they are correlated \cite{Sas2019}).

The PageRank of the cycles\footnote{Calculated as the maximum PageRank of the affected artefacts and normalized by the number of artefacts in a version.} is decreasing in 70\% of the instances, contrary to what we found in a previous study on open source Java systems \cite{Sas2019}.
The remaining 21\% of instances increase in PageRank, and only 9\% stay constant.

\begin{figure}
    \centering
    \includegraphics[width=0.5\textwidth]{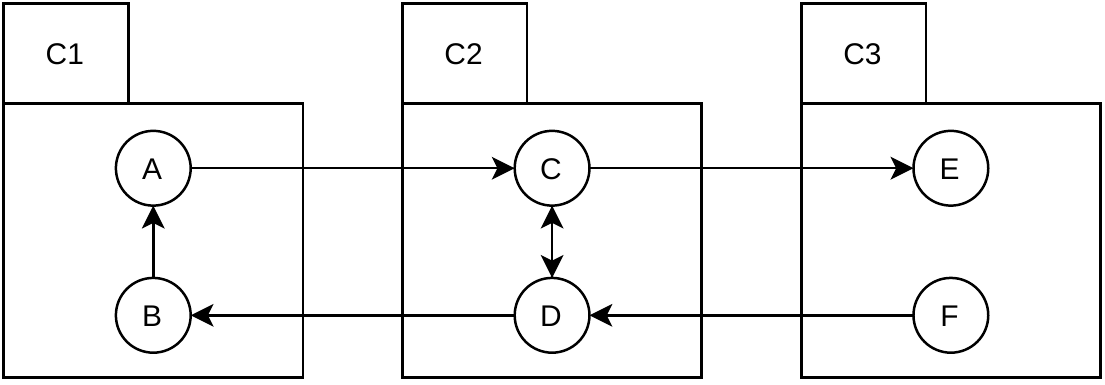}
    \caption{Example of two cycles among components: one among $C1$ and $C2$ that is also present among the files contained in them; and one among $C1$, $C2$ and $C3$ that is only present among the components. Figure adapted from \cite{AlMutawa2014}.}
    \label{fig:package-cycles}
\end{figure}

Typically, a file can have two types of dependencies (internal to the system), the first is to another file in the same component, and the second is to a file in another component.
Dependencies that cross the border of the component can also create cycles among components, either a) directly as a result of two or more files from the affected components creating a cycle among them; or b) indirectly, as a result of files that depend on files in another component but do not create a cycle among them, yet they create the dependencies among the components that in turn create the cycle (see Figure \ref{fig:package-cycles}). We call this characteristic `Affected design level' (see Table \ref{tab:characteristics}), and we used this characteristic to study how many cycles cross this border.
In the systems we analysed, 98\% of cycles are only among files, whereas the remaining 2\% are at the component level.
This means that, the vast majority of cycles is fully enclosed within the component their files belong to (i.e. they do not cross the component's border), which is a good sign of encapsulation but also means that \textbf{components are quite entangled internally}.
This could probably be because of the specific architecture of the system, which is divided in components that hide all the functionality under an interface.

\begin{figure}
    \begin{subfigure}[t]{0.65\textwidth}
        \centering
        \includegraphics[clip, trim=0cm 3.2cm 0cm 3.8cm,width=0.7\textwidth]{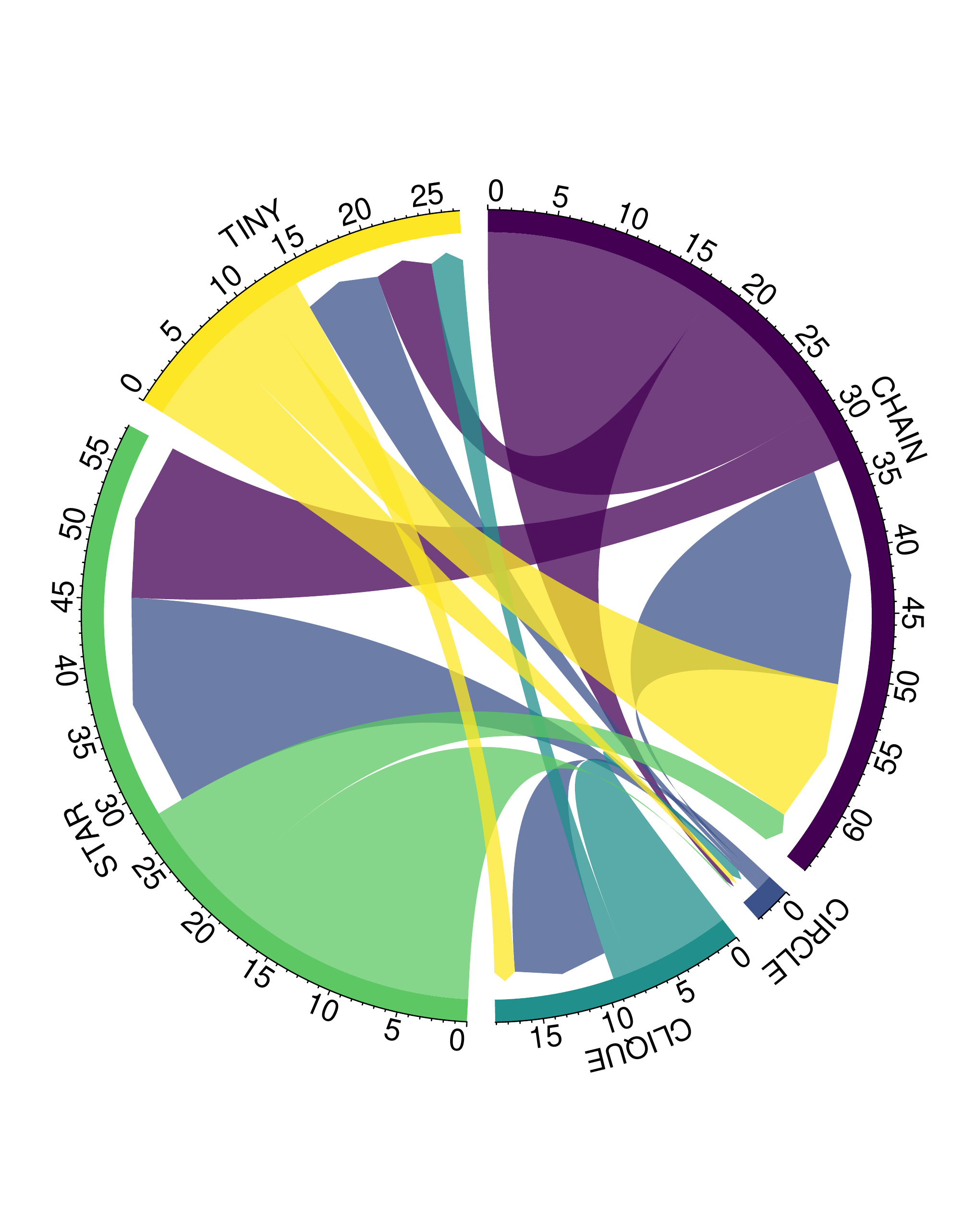}
        \caption{Chord diagram of how cycle shapes change, controlled for the distribution of shapes in the cycles population. The numbers represent the percentage of the total population of the corresponding shape.}\label{fig:chord-shapes}
    \end{subfigure}
    \hfill
    \begin{subfigure}[t]{0.3\textwidth}
        \centering
        \includegraphics[width=\textwidth]{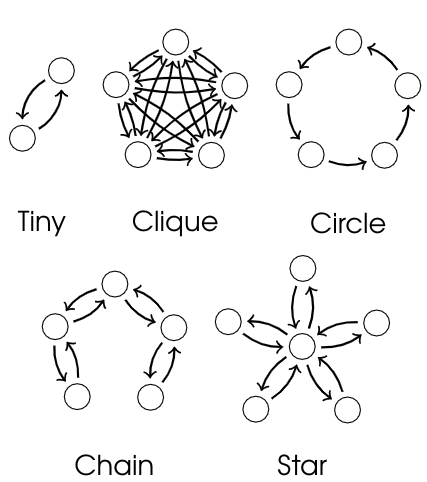}
        \caption{Symmetric cycle shapes detected by \textsc{Arcan} and defined by Al Mutawa et al. \cite{AlMutawa2014}.}\label{fig:cycle-shapes}
    \end{subfigure}
    \caption{The cycle shapes considered in this study and how they change over time. The total  population of cycles is as follows: Circle 86\%, Clique 7\%, Tiny 4\%, Chain 1.5\%, Star 0.5\%). Only instances that persists for at least 3 releases are considered.}
\end{figure}

Concerning the shape of the cycles, 73\% of the instances exhibit no change in shape over time, whereas the remaining ones (4206 instances) mutate as illustrated in Figure \ref{fig:chord-shapes}.
The chord diagram depicts the proportion of the cycle shapes that changed into another shape. Each sector of the diagram corresponds to a specific shape with outgoing edges that represent the proportion of the population of that shape that transforms into another shape. For example, only a tiny percentage of circle instances change shape, and therefore the corresponding sector of the circle shape is rather small, despite constituting the majority of the population of cycles (86\%).
As it can be noted, some shapes (i.e. chain and star) are more prone to changes than others, i.e. they have a greater percentage of their population changing.
There is also a certain balance across all shapes in the number of instances changing \emph{into} a shape and changing \emph{from} a shape. By looking more closely at the data, we notice that this was due to the fact that most instances bounce back and forth from one shape to the other.
The circle is a special case: despite only 5\% of circle instances being involved in changes, due to the sheer number of circle instances, the majority of changes involve circle shapes. Thus, circles are more likely to transform into any other shape, unlike stars for instance, which are more likely to change into chain or circle only.

\paragraph{Unstable Dependencies (UD)} For this smell type, as shown in Figure \ref{fig:dtw-classification}, 49\% of the 273 instances tend to remain constant in size over time, in 37\% of the cases UD increase in some way and the remaining 14\% of the times they experience a decrease of some sort. A similar behaviour is observed for the number of edges as well.

The PageRank of UD differs quite a lot from the other smell types, as we observe that 11\% of instances stay constant, 42\% have an increase of some sort, and 46\% experience a mostly gradual decrease. 
For other smell types, PageRank is mainly decreasing, whereas for unstable dependencies, a significant amount of instances exhibit an increase, meaning that they move towards more central parts of the system.
This means that components that are prone to change move towards more inner parts of the system.
This is not an ideal scenario, as Martin \cite{Martin2017} states that it is preferred to have dependencies that point toward more stable components in order to reduce change propagation.

Moreover, the gap in instability between the affected component and its dependencies is showing an increase in 38\% of instances, a decrease in 39\% of instances, while the remaining 22\% exhibit a constant trend. 
This means that there is no clear trend of instances that exhibit a clear increase, or decrease, in the instability measured in the central component and in its less stable dependencies (see Figure \ref{fig:ud} for more context).

The ratio of dependencies of an UD-affected component that are less stable than the affected component (i.e. the strength characteristic) was found to increase or decrease in equal percentages (28\% each), and stay constant in the remaining of cases (42\%).

\paragraph{Hublike Dependency (HL)} Hubs tend to either stay constant in size (49\% of instances) or increase (40\% of instances), with the remaining 11\% decreases.
Therefore, over time, hubs involve more and more artefacts.

By looking at the number of files within the central component that provide functionality to external components (incoming dependencies, i.e \emph{afferent ratio}\footnote{This is a ratio characteristic, however, for this analysis, it was weighted with the number of elements in the central artefact in the respective version to ensure we detected the absolute variations inside the internal artefact.\label{fn:ratio-characteristic}}) and at the ratio of files within the central component that use external components (outgoing dependencies, i.e. \emph{efferent ratio}\footref{fn:ratio-characteristic}) we note the following: the afferent ratio is increasing in 46\% of instances and decreasing in 11\% of instances only (remaining 43\% are constant); the efferent ratio on the other hand, exhibits an increase in 32\% of instances and a decrease in 36\% of instances (remaining 32\% are constant).
This means that at least \emph{some} HL instances tend to provide more functionality over time themselves rather than depending on their outgoing dependencies to provide such functionality.
This phenomenon is not optimal for the overall architecture of the system as it means that hubs, over time, \emph{replace} the functionality of their dependants: instead of having other dedicated components to provide that functionality, hubs take their place (i.e. they accumulate features).
The final result of this process is that hubs drift away from their initial purpose and become aggregators of functionality, weakening the separation of concerns originally intended by the architects. 

Finally, we observe the trend of the \emph{affected ratio}\footref{fn:ratio-characteristic}, i.e. the number of files within a hublike component that create the incoming and outgoing dependencies, thus creating the smell. This is increasing in 46\% of the cases, decreasing in 29\% of cases, and the remaining 25\% are constant.
Thus, as aforementioned, hubs grow to become more complex over time and more connected to their incoming and outgoing dependants.

\paragraph{God Component (GC)} The number of elements in the components affected by GC (i.e. size) increases in 53\% of the cases, stays constant in 40\% of the cases and decreases in 6\% of the cases.
Similarly, also the lines of code density increases 46\% of times, decreases in 34\%, and stays constant in the remaining 20\%.
We can therefore conclude that GC tend to grow in size over time, possibly aggregating more concerns and growing in complexity.

The PageRank of GCs follows a similar pattern as for the other smells, with 65\% of instances exhibiting a steady decrease, 24\% of them an increase, and the rest of them (11\%) stay constant.
This is a rather unexpected result as GCs, being large components by definition, are expected to also have an increase in their centrality over time.
This result however hints that the new functionality added in other parts of the system is ultimately less and less connected to the functionality offered by GCs given the decreasing PageRank of the majority of GC-affected components.
Such a pattern however is only observed globally in the whole dependency network of the system; locally, GCs still experience a growth in the number of files within the component and number of dependencies among those files (as mentioned above).

\paragraph{Summary of RQ1.1 results}
The general trend that we notice across the evolution of the smell characteristics is that each characteristic fits one of two patterns: it either (1) exhibits a \emph{dominant constant} trend followed by either an increasing or decreasing trend; or (2) it exhibits a \emph{dominant increasing} or \emph{decreasing} trend. 
The first case entails that those smell characteristics are mostly unaffected by the evolution of the smell. Examples of this case are CD Size, UD Size and CD Number of Edges.
In the second case, the opposite is true and the evolution of smell characteristics has a clear direction over time. Example of this case are PageRank for all smell types or GC Size.
This information can be exploited by using the smell characteristics of the second type as predictors for the evolution of an instance to establish the severity of a smell.
Instances with smell characteristics that have a clear trend and are bound to reach certain thresholds could be brought to the attention of developers before they become problematic and pose a greater threat to the maintainability and evolvability of the system.

\subsection{RQ1.2 -- Persistence of Architectural Smells in the system}
\subsubsection{Data analysis methodology: Survival analysis}
Different architectural smell types were found to have drastically different persistence rates within Java Open Source Systems \cite{Sas2019}.
To establish the persistence rates in our case (embedded systems written in C/C++), we employed the same technique used in our previous work \cite{Sas2019}: the Kaplan-Meier estimator, or survival analysis.
This technique is typically used in the biomedical sciences and in product reliability assessment; in addition, prior to our previous work \cite{Sas2019}, it was also employed in software engineering to analyse code smell persistence \cite{Chatzigeorgiou2014}.

Unlike simple descriptive statistics, such as mean or density functions, survival analysis also takes into consideration the possibility that a smell continues to affect the system even after the last version included in the analysis.
In the biomedical domain, this event is associated with the patient surviving past the period of the analysis.
More technically, this type of data is said to be \emph{right-censored}, because the outcome of the treatment cannot be measured, due to the conclusion of the study.

The survival analysis is performed using the Kaplan-Meier estimator \cite{Kaplan1958}, a non-parametric statistic that estimates the survival probability of a type of smell as the system evolves (new versions are released).
The statistic gives the probability $p$ that an individual patient (i.e. smell in our case), will survive past a particular time $t$.
At $t = 0$, the Kaplan-Meier estimator is equal to 1, and as $t$ goes to infinity, the estimator goes to 0. Also, the probability of surviving past a certain point $t$ is equal to the product of the observed survival rates until $t$.

\subsubsection{Results}
\begin{figure}
    \begin{subfigure}[b]{0.49\textwidth}
        \centering
        \includegraphics[width=\textwidth]{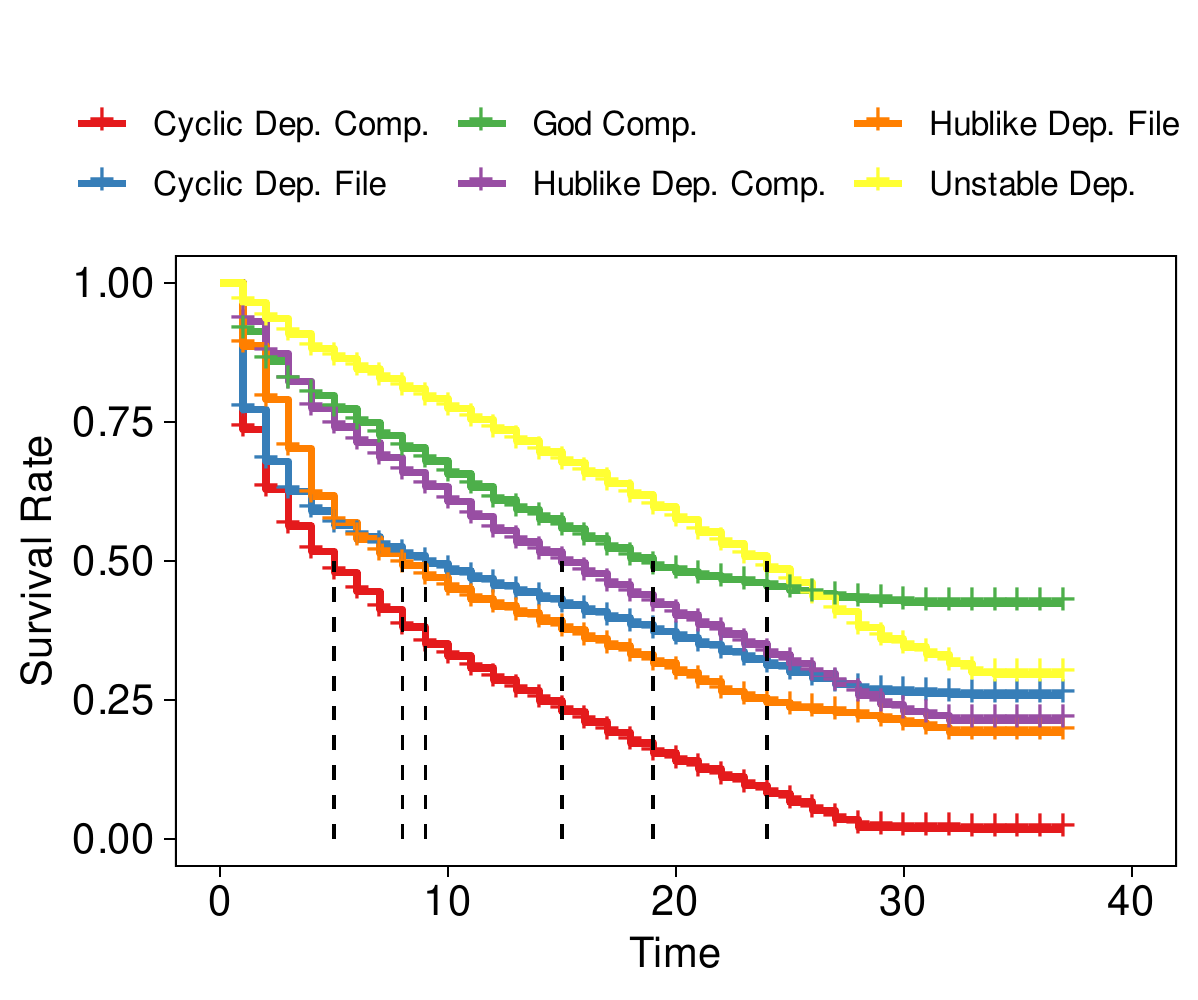}
        \caption{Survival rate of smell types.}\label{fig:survival-smells}
    \end{subfigure}
    \hfill
    \begin{subfigure}[b]{0.49\textwidth}
        \centering
        \includegraphics[width=\textwidth]{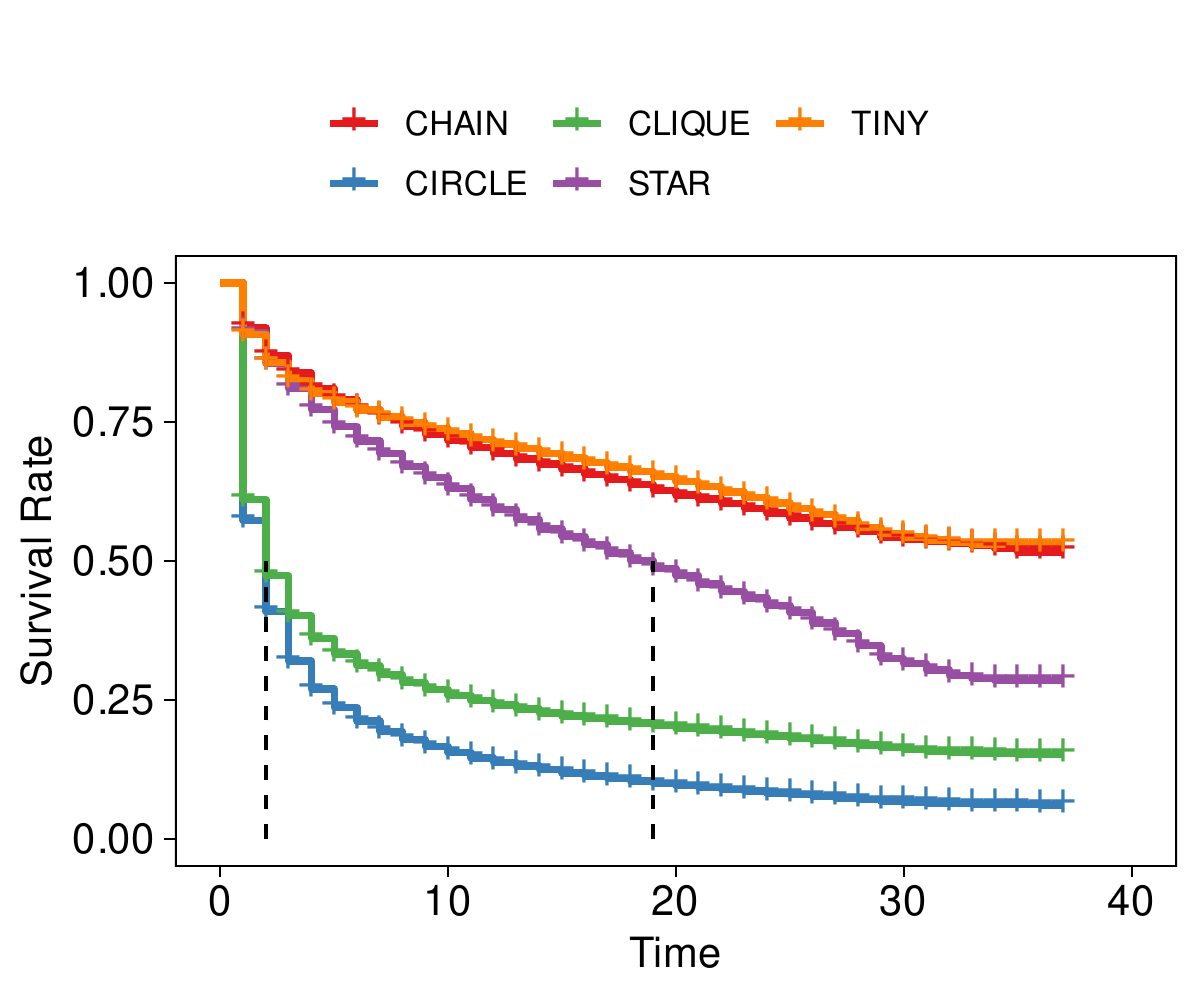}
        \caption{Survival rate of cycle shapes.}\label{fig:survival-shapes}
    \end{subfigure}
    \caption{A visualisation of the Kaplan-Meier estimators. The plot reads as follows: after a certain time $t$ (on the $x$ axis), smell type $s$ has a probability $p$ (on the $y$ axis) to survive. Dashed vertical lines represent the value $t$ when $p = .5$.}
    \label{fig:survival-analysis}
\end{figure}

The results of this analysis are presented in Figure \ref{fig:survival-analysis}. The figure shows the survival rate for both smell types and cycle shapes.
Figure \ref{fig:survival-smells} differentiates between smells at file and component level for cycles and hubs: the appearance and disappearance rates of dependencies among files and dependencies among components may be different, thus we study them separately.
Given their definitions, UDs and GCs cannot be detected at file level; therefore, we only considered them at component level.

\paragraph{Smell types}
By looking at Figure \ref{fig:survival-smells}, one can note that the smell type with the lowest survival rate are cyclic dependencies among components, which tend to disappear from the system rather quickly: they exhibit a 50\% probability of surviving more than 6 versions. 
Cycles at file level instead manage to affect the system for a little bit longer, reaching 50\% probability of surviving after 9 versions. This makes sense as it is much more likely for developers (in the company subject to this study) to eliminate unwanted dependencies towards files in external components, rather than towards internal files.
Hubs show a similar survival rate and reach the 50\% probability of surviving at 8 versions, at file level, and 16 versions at component level, before converging later on.
God components and Unstable dependencies reach it at 19 and 24 versions, respectively.
God components, however, maintain a flatter curve and stay close to the 50\% threshold for longer.
Another interesting fact that can be derived from Figure \ref{fig:survival-smells} is that the curves stabilise eventually (see the right-most part of the plot) and do not go below a certain probability (excluding cycles at component level).
This is probably due to the fact that the parts of the system affected by smells for a long time tend to become legacy code that is either very hard to change or has no reason to be changed. Our interviews have provided some insights into this phenomenon, which we will explore in more depth in the discussion section (Section \ref{sec:discussion}).

\paragraph{Cycle shapes}
In Figure \ref{fig:survival-shapes}, we focus on the survival rates of cyclic dependencies, regardless of the type of artefact they affect, and distinguish between different shapes.
Circles are the ones that are more likely to disappear from the system (50\% chance of surviving for one version), however, they are also the most common type of shape and much easier to form, especially in comparison with chain, clique and star.
Cliques, despite being a much more complex type of shape, have a similar survival rate to the one we observed for circles.
This is probably due to the fact that cliques are less common, harder to appear, and can be ``broken'' just by removing one edge from their structure.
Moving to stars, despite being relatively complex (and thus relatively easy to break down), they manage to survive within the system for a much longer time, reaching 50\% of survival probability only after 17 versions.
Finally, chain and tiny shapes are the ones that exhibit the longer survival rate while also having a relatively stable curve.
This is probably because: a) these shapes are very similar; b) cycles between fewer elements are less likely to be perceived as problematic - in fact, they could be intentional.

\section{RQ2 -- Architectural smells co-occurrence}\label{sec:rq2}
\subsection{Data analysis methodology}\label{sec:methodology-rq2}
To find out what pairs of architectural smells co-occur more often, we used a simple approach: we calculated the co-occurrence matrix for each type of architectural smell detected by \textsc{Arcan}. This resulted in a $6\times6$ matrix, where the rows and columns are labelled with the names of the smells. However, for the sake of readability, we report the results in two matrices, one $4\times4$ matrix for component-level smells and one $2\times2$ matrix for file-level smells. 
The value in each cell of these two matrices is calculated as follows: 
\begin{equation}
    cooc_{i,j} = \frac{\# \textnormal{ of instances of type } i \textnormal{ overlapping one of type } j}{\# \textnormal{ of total instances of type } i} \times 100
\end{equation}
with $i \neq j$.
By `overlapping' we mean that the two smell instances must affect at least one artefact in common in the same version.
However, some architectural smells involve various artefacts which play different roles; thus we also distinguish between the different parts of the smell that may overlap:
\begin{itemize}
    \item for Hublike Dependencies we distinguished between the incoming dependencies (artefacts C1-3 in Figure \ref{fig:hl}), outgoing dependencies (artefacts B1-3 in Figure \ref{fig:hl}), and the central component, or the hub (artefact A in Figure \ref{fig:hl});
    \item for Unstable Dependencies we distinguished between the central component (component A in Figure \ref{fig:ud}) and its outgoing dependencies that are less stable (components B1-3 in Figure \ref{fig:ud});
    \item for Cyclic Dependencies we did not make any distinction, as every component of the cycle plays a similar role in the smell;
    \item for God Component we did not make any distinction as the smell constitutes a single element.
\end{itemize}

Note that for this analysis, we counted every smell detected individually, \emph{without} linking it to its corresponding instances in adjacent versions. This way, we capture not only the overlaps that take place in multiple versions but also those that happen in one version; thus we represent a more precise picture of the overlaps of smells. Moreover, this approach is very similar to what was done in a previous study on code smells \cite{Palomba2018}.

\subsection{Results}\label{sec:results-rq2}
The results obtained for this research question are reported in Table \ref{tab:co-occurrence}, for component-level smells, and in Table \ref{tab:co-occurrences-files} for file-level smells.
The values in the table represent the percentage of the total number of instances of the smell in the corresponding row that overlap with the smell in the corresponding column (hence the table is not symmetrical).

\paragraph{Component-level smells}
With a first glance at Table \ref{tab:co-occurrence}, one can note that the architectural smells in the analysed systems have a very high overlap, which is reasonable given the definition of some smells (i.e. they involve numerous components).

Looking at the CDs in Table \ref{tab:co-occurrence}, we note that given their abundant presence in the system, they overlap with the other smell types in high percentages (from 76\% to 99\%, as seen in the first row). 
This is most likely due to the fact that cycles affect multiple elements, and its easier for an instance to overlap with another instance of a different type.
Nonetheless, it is interesting to note a discrepancy between how many CD instances overlap with a GC (86\%), and how many GC instances overlap with a CD (58\%). This is because several god components take part in multiple cycles: a significant number of cycles (86\% of 12135) overlap with a GC but there are only 3165 instances of GC, which means that \textbf{multiple cycles must be affecting the same GC instances}. 

Concerning HL instances, it is interesting to note that 74\% of hubs (centres) are also unstable, meaning that the risk of changes propagating to their dependants is increased. 
We also note that hubs can be intentional design choices that expose low-level functionality to components with a high level of abstraction under a single interface (as mentioned by some interviewees).
Nonetheless, this could be a double-edged sword: while hubs might serve the purpose of abstracting low-level functionality, they might also increase the likelihood of changes propagating from low-level components to unrelated high-level components.
In addition, as Martin mentions (see the Stable Dependencies Principle \cite{Martin2017}) this could also mean that they \textbf{become harder to change, because there is a lot of high-level functionality that \emph{might} depend} on it but it is hidden to developers by the central hub.

God Component, compared to the other smell types, exhibits fewer overlaps.
This low interaction rate is particularly notable with hubs, as only 10\% of GCs are also hubs (centres of HL).
This highlights how the two smell types centralise logic differently: \emph{GCs aggregate implementation}, and therefore they grow in number of lines of code, whereas \emph{HLs aggregate abstractions and delegation}, and therefore they grow in number of incoming and outgoing dependencies.
Furthermore, we observe that 46\% of GC instances are also UD instances whereas we see only 29\% in the opposite case. This means that \textbf{46\% of the GC instances}, which aggregate functionality and thus increase in size and complexity, \textbf{are more likely to change due to changes in \emph{neighbouring} components}.

Unstable Dependencies were mostly covered when discussing the other smell types, but it is still noteworthy to mention that 52\% of them have their centre taking part in a cycle and 97\% of all cycles go through an unstable dependency centre.
This \textbf{increases the chance of changes propagating} to other components and ripple through the elements affected by the cycle.
Moreover, we note that only 8\% of UDs are hubs, which makes sense as the definition of UD is not based on the number of incoming/outgoing dependencies (unlike HL); this means that it can be detected in more parts of the system, thus explaining the small percentage of overlaps.

\paragraph{File-level smells}
Looking at Table \ref{tab:co-occurrences-files} we note that the number of cycles among files and the number of hubs among files differ by two orders of magnitude.
However, we still observe that a lot of cycles (14\%) have an overlap with hubs at file level, which means that one or more cycles go through a hub.
Likewise, 94\% of hubs, 97\% of incoming and 99\% of outgoing dependencies are also involved in cycles.

The high number of cycles and their overlap with hubs suggests that the dependencies internal to the components are tightly coupled. This makes changes hard to implement, because it may not be clear how responsibilities are shared between files and how a change will impact other files.
This means that hubs at file-level are a very likely to be a \textbf{maintenance hotspot}, as they not only accumulate responsibilities, but they are also a sign of high coupling among the hub, the files depending on it, and the files it depends upon caused by the cycles among those very files.
We caution, however, that these may only be specific to the projects analysed and not applicable in a different context.

\begin{table}[tbp]
    \footnotesize
    \centering
    \caption{Co-occurrence (or overlap) of component-level architectural smell types. 
    Percentages refer to the total number of instances, shown in the right-most column. Key values are underlined and in bold face.}
    \label{tab:co-occurrence}
    \begin{tabular}{@{}m{0.35cm}|r|cm{0.65cm}m{0.65cm}cccc|c@{}}
    \toprule
    \multicolumn{2}{r|}{\multirow{2}{*}{\textbf{Smell Type}}} & \multicolumn{1}{c|}{\multirow{2}{*}{\textbf{CD}}} & \multicolumn{2}{c|}{\textbf{UD}} & \multicolumn{3}{c|}{\textbf{HL}} & \multirow{2}{*}{\textbf{GC}} & \multirow{2}{*}{\textbf{\begin{tabular}[c]{@{}c@{}}Total\\ Instances\end{tabular}}} \\ \cmidrule(lr){4-8}
    \multicolumn{2}{r|}{} & \multicolumn{1}{c|}{} & \multicolumn{1}{c}{less stable} & \multicolumn{1}{l|}{\textbf{centre}} & \multicolumn{1}{l}{incoming} & \multicolumn{1}{l}{\textbf{centre}} & \multicolumn{1}{l|}{outgoing} &  &  \\ \midrule
    \multicolumn{2}{r|}{\textbf{CD}} & - & 99 \% & 97 \% & 91 \% & 76 \% & 94 \% & \underline{\textbf{86}} \% & 12135 \\ \cmidrule(r){1-2}
    \multirow{2}{*}{\textbf{UD}} & less stable & 92 \% & - & 83 \% & 61 \% & 42 \% & 83 \% & 59 \% & \multirow{2}{*}{5121} \\
     & \textbf{centre} & \underline{\textbf{52 \%}} & 50 \% & - & 49 \% & \underline{\textbf{8 \%}} & 32 \% & \underline{\textbf{29 \%}} &  \\ \cmidrule(r){1-2}
    \multirow{3}{*}{\textbf{HL}} & incoming & 89 \% & 93 \% & 94 \% & - & 55 \% & 77 \% & 74 \% & \multirow{3}{*}{587} \\
     & \textbf{centre} & \underline{\textbf{77 \%}} & 82 \% & \underline{\textbf{74 \%}} & 53 \% & - & 55 \% & \underline{\textbf{59 \%}} &  \\
     & outgoing & 95 \% & 100\% & 90 \% & 79 \% & 53 \% & - & 78 \% &  \\ \cmidrule(r){1-2}
    \multicolumn{2}{r|}{\textbf{GC}} & \underline{\textbf{58 \%}} & 60 \% & \underline{\textbf{46 \%}} & 41 \% & \underline{\textbf{10 \%}} & 42 \% & - & 3165 \\ \bottomrule
    \end{tabular}\\
    \vspace{3mm}
    {\raggedright \textbf{CD}: Cyclic Dep. \textbf{HL}: Hublike Dep.; \textbf{UD}: Unstable Dep.; \textbf{GC}: God Comp. }
\end{table}

\begin{table}[tbp]
    \footnotesize
    \centering
    \caption{Co-occurrences (or overlap) of file-level architectural smell types. 
    Percentages refer to the total number of instances, shown in the right-most column. Key values are underlined and in bold face.}
    \label{tab:co-occurrences-files}
    \begin{tabular}{r|r|cccc|c}
    \toprule
    \multicolumn{2}{r|}{\multirow{2}{*}{\textbf{Smell Type}}} & \multirow{2}{*}{\textbf{CD}} & \multicolumn{3}{c|}{\textbf{HL}} & \multirow{2}{*}{\textbf{\begin{tabular}[c]{@{}c@{}}Total\\ Instances\end{tabular}}} \\ \cline{4-6} 
    \multicolumn{2}{r|}{} &  & incoming & \textbf{centre} & outgoing & \\ \midrule
    \multicolumn{2}{r|}{\textbf{CD}} & - & 27 \% & \underline{\textbf{14 \%}} & 44 \% & 203646\\ \cmidrule{1-2}
    \multirow{3}{*}{\textbf{HL}} & incoming & 97 \% & - & 54 \% & 88 \% & \multirow{3}{*}{1345} \\
     & \textbf{centre} & \underline{\textbf{94 \%}} & 55 \% & - & 54 \% \\
     & outgoing & 99 \% & 91 \% & 55 \% & - \\ \bottomrule
    \end{tabular} \\  
    \vspace{3mm}
    {\raggedright \textbf{CD}: Cyclic Dep. \textbf{HL}: Hublike Dep.}
\end{table}

\section{RQ3 -- Architectural smells precedence}\label{sec:rq-3}
\subsection{Methodology}
Similarly to the previous RQ, to calculate the number of times a smell type is introduced before another smell type, we used a matrix.
For each architectural smell type $i$ and $j$ (with $i \neq j$):

\begin{equation}
    intr^k_{i,j} = \frac{\# \textnormal{ of times an instance of type } i \textnormal{ preceded one of type } j}{\# \textnormal{ times AS instance of types } i \textnormal{ and } j \textnormal{ overlap within } k \textnormal{ versions}} \times 100
\end{equation}

To obtain more insight, we look into how many versions it usually takes for a smell of a different type to be introduced. To this end, we repeated the calculation by counting the times that a smell type $i$ was introduced before another smell type $j$ if and only if $j$ was introduced at max $k$ versions after $i$, with $1\le k \le 37$. In total, we ended up with 37 matrices, one matrix for each value of $k$. Note that $37$ was chosen because it is the maximum number of versions we analysed. 
This setting allows us to understand how the precedence values vary when looking farther in time (i.e. larger values of $k$).

\subsection{Results}
\begin{figure}
    \centering
    \includegraphics[width=0.95\textwidth]{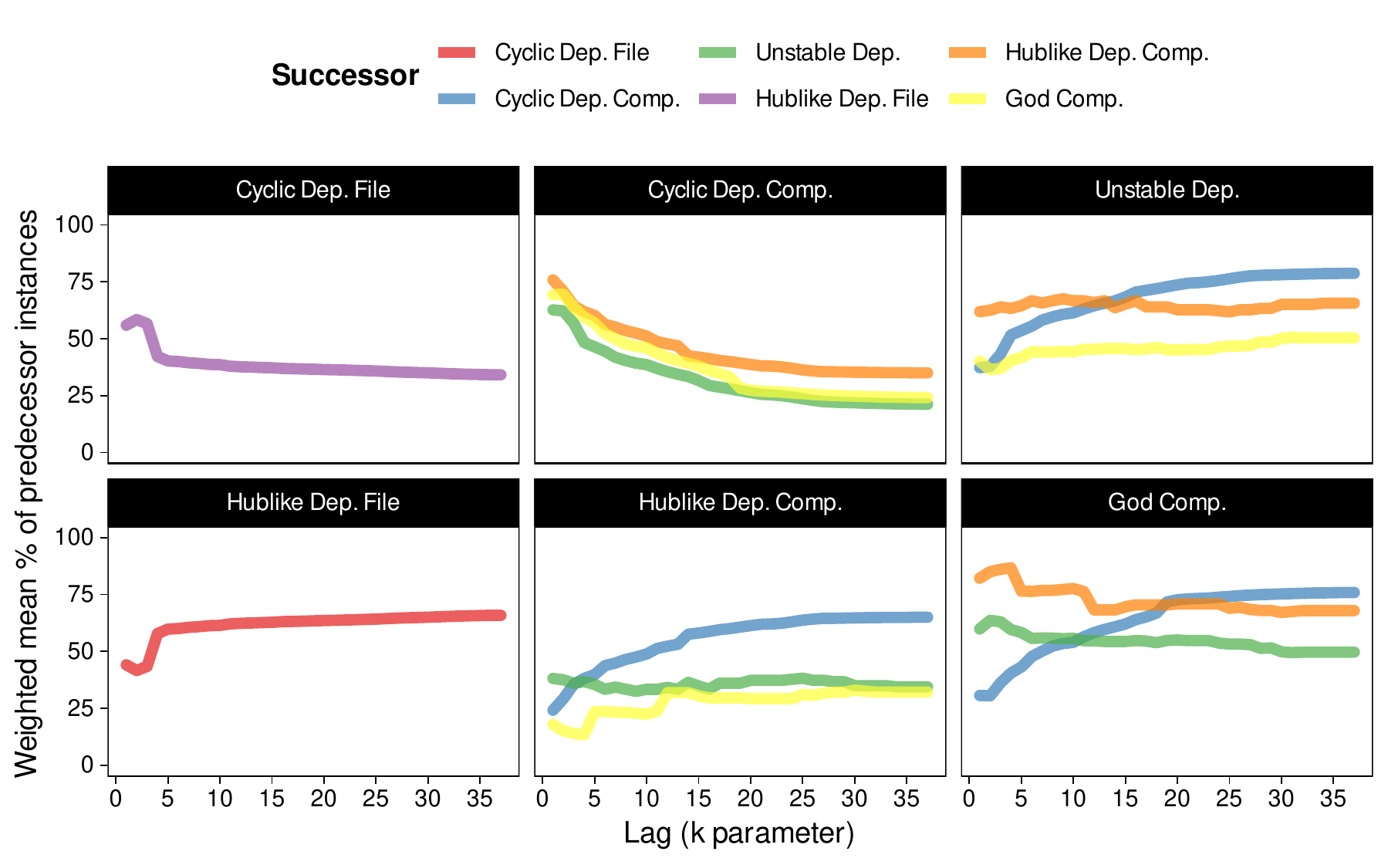}
    \caption{The percentage of instances for each smell type that precede the other smell types, measured for different values of $k$. Each quadrant represent the predecessor smell type. Percentages are weighted by number of occurrences in each project for a given value of $k$.}\label{fig:precedence}
\end{figure}

The results for this research question are presented in Figure \ref{fig:precedence}.
The figure shows the values assumed by $intr_{i,j}$ for different values of $k$. Each quadrant shows the percentages of instances where the smell type $i$ is the predecessor of an instance of smell type $j$ in percentage of the number of times instances of type $i$ and $j$ overlapped within $k$ versions.

CD instances tend to precede the other smell instances by one release ($k = 1$) in 60\% to 80\% of the cases, depending on the smell. For small values of $k$, file-level cycles precede hubs in more than 50\% of cases; whereas for $k = 37$, this is less likely to happen as cycles have rather short lifespans (see RQ1.2 results), so the percentages plunge down to 30\%.
Component-level cycles, instead, precede the introduction of other smell types rather commonly, reaching up to 75\% for $k = 1$, meaning that \textbf{as soon a cycle appears} it is very likely that \textbf{another smell will affect one of the components in the cycle}.
Similarly to file-level cycles, component-level cycles also have a short lifespan, so the percentages of precedence follow the same pattern.

For small values of $k$, UD instances are likely to precede HL instances in the same component (60\% of the cases), with GC and CD being a bit less likely.
Since CD instances are much more common, when using higher values of $k$, they are much more likely to succeed UD instances (75\%).
These results hint that the frequent changes affecting UD instances are very likely to result in UD instances will overlap with a CD, GC, or HL down the road, possibly due to the higher instability of their dependencies that force them to change more often and develop other smells.

GC instances seem to have the highest variability, with 75\% of instances preceding HLs, 55\% preceding UDs, and 30\% preceding CDs.
This means that the \textbf{complexity of a GC is very likely to introduce other smell instances} such as a HL and/or a UD.
Only when $k$ is larger, a CD instance is eventually introduced.

HL instances at the component-level, on the other hand, are much less likely to precede another instance, especially on the short term ($k \le 3$). UD instances are the most likely at 35\%, followed by CD at 25 \% and GC at 23\%.
File-level HL instances are likely to precede CD (almost 50\% of HLs do so) because CD are ubiquitous.
However, what is most interesting, is when we consider how HL ranked in the results of other smell types. 
We note that HL are usually more likely to appear after other smell types, in fact they are always \textbf{the most likely smell type to appear after a smell of another type was introduced}.

\section{RQ4 and RQ5 -- Practitioners and Architectural Smells}\label{sec:rq-4-and-5}
\subsection{Data analysis methodology}\label{sec:methodology-rq-4-and-5}
The qualitative analysis adopted the Constant Comparative Method (CCM) \cite{Glaser2017, Boeije2002}, part of Grounded Theory \cite{Glaser1968}, to deduct valuable insights from the interviews. Grounded Theory (GT) is one of the most important methods in the field of qualitative data analysis. It has been used extensively within both social sciences and software engineering and provides a structured approach to process and analyse the data collected from multiple sources. GT increases the theoretical sensitivity of the researcher as the data analysis progresses and eventually allows to formulate hypotheses and theory \cite{Glaser1968}.

As mentioned above, we have used CCM, an inductive data coding and categorization process that allows a unit of data (e.g., interview transcript, observation, document) to be analyzed and broken into codes based on emerging themes and concepts; these are then organized into categories that reflect an analytic understanding of the coded entities \cite{Mathison2005}.

\begin{figure}
    \centering
    \includegraphics[width=.8\linewidth]{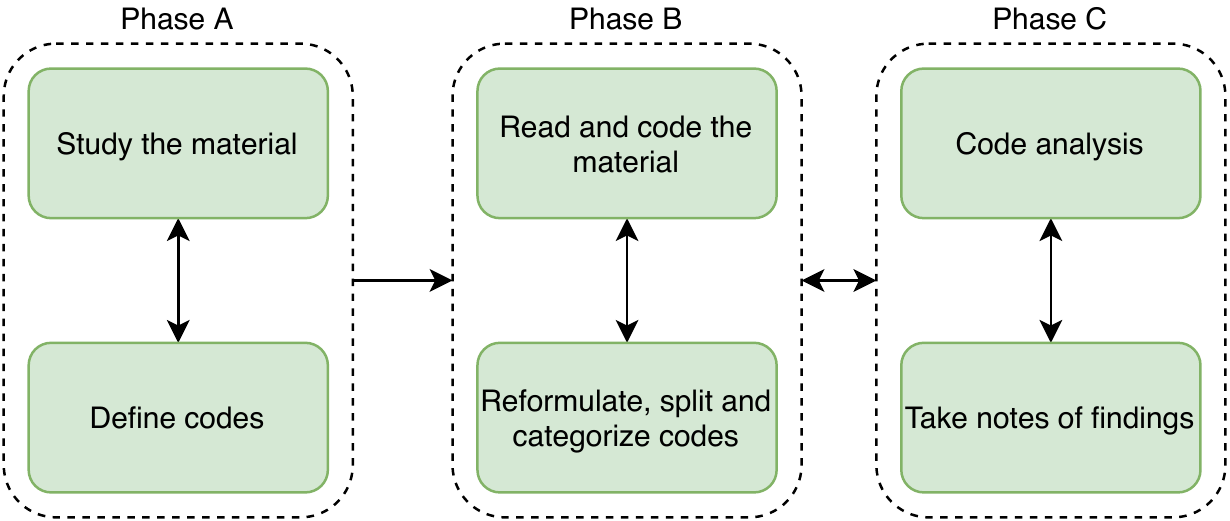}
    \caption{The phases of the qualitative data analysis process.}\label{fig:qualitative-analysis}
\end{figure}

The qualitative data analysis process is presented in Figure \ref{fig:qualitative-analysis}. During the first phase (Phase A), the collected material (i.e. interview recordings) was studied and a code map was created to organise the codes used to tag the data.
After completing this phase, the coding process started (Phase B), which also involved updating and re-organising the codes based on the new understanding of the data.
As new interviews were recorded, transcribed, and coded, the data was also gradually analysed and notes were taken with the aid of the codes in the data (Phase C).
To aid with the organisation of the codes, we created a network of codes\footnote{See replication package.}, where each code was linked to other codes based on their relationship.
In total, two rounds of coding where done, the first one as interviews were transcribed, and the second one after the transcribing process was completed, to ensure that the codes added along the way were present in all the data.
Additionally, coded quotations from the interviews that referred to the same topic (e.g. two participants referring to the same event) were linked together to help navigate the quotations during data analysis. This process included both intra- and inter-document quotations, where documents refer to interview transcripts.
The whole process was performed by the first author of the paper, while the second author reviewed the codes and coding schemes as they were developed to reduce the risk of biases (e.g. confirmation and information bias).
To automate the data analysis as much as possible, we relied on Atlas.ti \footnote{See \url{https://atlasti.com/}.}, a dedicated qualitative data analysis tool.

\subsection{Results}\label{sec:results-rq-4-and-5}

\subsubsection{RQ4 -- Support to practitioners}\label{sec:results-rq-4}
\paragraph{Overall considerations}
Most of the interviewed participants stated that the reported results resembled what their intuition and expectations were prior to seeing the report.
\begin{quote}
    \emph{``It was more like a confirmation, because yeah, since I was busy with this project for five years, I had a feeling where the “bottlenecks” were and which components were changed the most."}
\end{quote}
Many practitioners also reported that the results correlate with the parts of the system they experienced issues (either currently or in the past).
The most unexpected result for some participants was the number of Cyclic Dependencies affecting the files within a certain component; they mostly underestimated it, particularly for components that are relatively new.
\begin{quote}
   \emph{``Something that I didn’t knew is that Component X and Component Y are also not doing good while they are relatively new components. "}
\end{quote}

This begs the question whether the architectural smells analysis actually helps architects and developers, since they already know where the issues are. Participants mentioned that the report provides them with the following benefits: \emph{(a)} a ``good view" of \textbf{where cyclic dependencies are} so they do not have to ``grope in the dark", \emph{(b)} a way to \textbf{prioritise} the future improvements based on where exactly the current smells are, \emph{(c)} a good idea of \textbf{how complex and extended a specific change} (e.g. add/modify a requirement) could be, \emph{(d)} a way to \textbf{track the issues}, making them visible to the rest of the team, \emph{(e)} a clear approach to determine \textbf{when an issue has been fixed}, and \emph{(f)} a way to find out if the \textbf{problem reappears} in the future.

A common point among all these benefits is that they all contribute, in one way or another, to sharing the knowledge of the problems present in the project with all team members in a way that would otherwise be tacit.
One participant also highlighted the usefulness of the information provided for new team members:
\begin{quote}
    \emph{``[...] this will be very useful, for example, to any person coming to the team or a new architect of a team. Graphs like this will then provide years of experience in one go."}
\end{quote}

Transferring and tracking knowledge as a team can be rather cumbersome \cite{Rus2002}, so automating this task with a tool, is an added value that several practitioners appreciated, and expressed a desire to integrate into their workflow in order to receive periodical reports.

Finally, the fact that the reported issues are already known to practitioners, is considered as a positive outcome of our study. It indicates that the AS we were able to identify are true hotspots within the system (though quantifying this using the Precision and Recall metrics was out of the scope of this study).
It is also worth noting that in some cases, the problematic components highlighted in the results were already part of the quality improvement roadmap that one designer proposed to the architect responsible for their project. 

\paragraph{Specific feedback}
After having established that the analysis actually provides an added value to the practitioners, we now describe which details provided them with the most insights.

In terms of the information contained in the report that the participants marked as useful, or referenced while explaining something, or implied that it allowed them to plan future activities accordingly, we have the following:
\begin{itemize}
    \item the dependency graph of components, as it provided an overview of the current state of the system's architecture;
    \item the heatmap showing what components were affected by most smells (and what type these smells were), as it allowed to identify the hotspots of the systems quickly, and plan accordingly;
    \item the total number of smells (divided by type) over time, as it showed the trend of the quality of the system;
    \item the histogram with the number of incoming and outgoing dependencies for each component in the current version, as it shows an overview of the system like the dependency graph but allows for an easier comparison between the components;
    \item the number of components involved in a smell (i.e. the Size characteristic), and other characteristics (like shape of a cycle), as it provided a quick summary about the smell and its possible effect on the system in a glance;
\end{itemize}
The fact that all this information could be generated automatically, with very little configuration by the user, and on-demand, was greatly valued.

\paragraph{Missing information}
The participants also provided their opinion on what information is missing from the report.
A common feedback that we received is the lack of ability to dive into the details of a specific smell, and visualise the relationship between the affected components, how they interact with their neighbours, and other contextual information useful to fix that smell.
However, it is fair to note that this was also not the purpose of the report to begin with; rather, it was designed to provide a general overview of the architectural smells present in the system.

\subsubsection{RQ5 -- Impact on Maintainability and Evolvability}\label{sec:results-rq5}
During the interviews, practitioners shared several experiences concerning the maintainability and evolvability of components that were affected by smells.
Although the majority of these anecdotes referred to different events and projects, most of them had enough similarities to allow us to identify a few patterns in the types of issues faced when trying to maintain or evolve the system.

\paragraph{Ripple effects}
The most common type of problem is related to the \emph{ripple effects} of changes. Making any kind of change to components affected by smells, is a 
troublesome process that required additional effort to carry out.
This additional effort was mostly due to changes that would \emph{propagate} to parts of the system that were partially, or totally, unrelated to the original change.
\begin{quote}
    \emph{``When I consider our changes in the past, these components are almost always touched. Depending on whether we can keep a change internal to a component or not, it may be that the change propagates to interface of this component. If it does, then we get this domino effect.''}
\end{quote}
Change propagation (or change ripple effects) is problematic, as changing a component might propagate to different components belonging to different teams.
This often means that a simple change could impact multiple teams, thus requiring further synchronisation between the teams to get it done; ultimately, the change becomes much costlier.
The same participant that gave the previous quote, provided an example of this phenomenon after being asked whether he/she noticed a correlation between changes and the components affected by smells:
\begin{quote}
    \emph{``Two years ago I made some changes in Component X that propagated to 56 components only because we changed the interface of that component. The changes we made were big and not backwards compatible and we had to change almost 60 different components in 5-6 teams, and it took a year to get everything done.''}
\end{quote}
In this case, Component X was affected by both God Component and Unstable Dependency, and the subject also mentioned Unstable Dependency as the most critical type of smell before providing this example.
We cannot claim of course, that the presence of the two smells are directly the cause of the ripple effects of these changes in other components.
However, the answers provided by the subjects directly link the presence of smells with an increased change propagation and change-proneness in the affected and neighbouring components.

On a similar note, another practitioner mentioned an example where changes in the code belonging to low-layer components that control the hardware, propagated to components in higher layers, even though that was not supposed to happen.
The low-level components were responsible for controlling some underlying sensors and hardware with the goal to support the hardware of a new machine.
The changes to these components triggered changes that rippled upward in the hierarchy of layers and the amount of work required to complete the update process was \textbf{initially underestimated}. The subject linked this particular case to both Hublike Dependency and Unstable Dependency: the low-level hardware components were controlled by a middle-level component, which was both unstable and a hub, and the high-level components depended on it.

Finally, ripple effects were also commonly associated with god components and their inherent internal complexity as well as with the fact that they usually contained a lot of legacy code. Making changes to complex god components was considered a \emph{risk}, because every change could affect multiple files and change the behaviour in unknown parts of the system or component (as pointed out by one of the practitioners this is also due to the inadequacy of their tests to test for regression).
Some god components contained files that were so tangled (i.e. affected by cycles) that even a simple change would have impacted several other files.

\paragraph{Architecture Erosion}
In addition to changes rippling to external components, participants also provided examples that the presence of smells is a sign of \emph{architectural erosion} \cite{Perry1992}: the gap between the original, intended architecture and the actually implemented architecture, that happens due to the continuous maintenance and evolution activities.

One of the interviewed architects explains their struggle with implementing the parallelisation of two tasks in order to speed up the production throughput of the whole machine. 
The tasks were both implemented by a certain Component Y, which, over time, became so complex and intricate (and also contained legacy code) that made it too difficult to proceed with the implementation of the desired feature (i.e. parallelisation) before actually refactoring the code\footnote{Note that the refactored version of the code was implemented in another component, which, when it is ready, it will supersede Component Y.}.
\begin{quote}
    \emph{``In your results Component Y is both a god component and a hublike dependency. [...] We do have maintenance issues in that component, so we want to split it to smaller functions because it has a lot of functionality and is quite a drawback to scale the functionality [...]. The road map that we had for improving it is to split the component to allow us to do the two things in parallel.
    To make that first step it was very painful in the short term, but once we got the hang of it it’s been improving a little bit.''}
\end{quote}
This example reflects how an important evolution of the system that would provide a tangible improvement for the customer, is hindered by: a) the centralisation of functionality in a single component (i.e. the two tasks), which is typical of hubs; and b) the aggregation of implementation (and legacy code), which is typical of god components. This component was originally not meant to be so large and complex, but erosion happened over time.

Cyclic dependencies among components were also mentioned by multiple architects as a sign of architecture erosion.
One architect mentioned an interesting example of how cycles were creating, over time, various problems that confused the team about what responsibilities were implemented by what component.
\begin{quote}
    \emph{``We also had a famous cyclic dependency between Components Z, U, and V, which had all kind of interesting things. Over time, sometime one controlled the other and sometimes the other controlled the first. That always gave us problems. So we are now actively redesigning that part to get rid of that cycle.''}
\end{quote}
This is a textbook example of the detrimental effects of cyclic dependencies among components on the maintainability of the system. Since the original architecture is eroded, developers first have to reverse engineer the responsibilities of the components at that point in time before applying the desired change to the system.

Moreover, another architect provided a very interesting anecdote about trying to refactor one cyclic dependency, showing how hidden dependencies, and the resulting complexity can ultimately have a direct impact on the company's business.
\begin{quote}
    \emph{``Last year we tried to remove a cyclic dependency by introducing a pattern to remove part of the cycle. However, we were not aware of all the legacy functionality within that component, and while removing the cycle, we missed some of the dependencies. This led to a lot of escalation in the field and we had to fly over to our customer to explain why this happened. It was actually a combination of god component and cyclic dependency. It was a real pain and the whole team had to work for two or three months to get it solved.''}
\end{quote}

In conclusion, all the examples mentioned in this section highlight how certain smell types (i.e. god components, hubs, and cycles), in one way or another, hinder the evolution -- and even the refactoring -- of the system and reflect its architecture erosion, further preventing developers to deliver new functionality to the customer.

\paragraph{Bugs and errors}
Practitioners also shared stories on how certain smell types affect the correct functioning of the system.

Cyclic dependencies, for instance, were mentioned several times (by multiple subjects) as a type of smell that causes errors at runtime, such as deadlocks, synchronisation issues of two or more tasks working together, or reduced throughput.
\begin{quote}
    \emph{``For example, if we look at the cyclic dependencies in the report. And look at the first one you see, this is basically the interaction between dose control peripherals. Those [components] are basically sensors, and these peripherals should only talk to [a master component] without talking between each other. […] when you have time-critical data coming in, this can lead to some timing errors in the field and sometimes deadlock.''}
\end{quote}

God components were also mentioned when discussing bugs and errors, though cycles among components were more dreaded because they had a direct impact on the observed behaviour of the system by the customer.

\paragraph{Communication}
Finally, practitioners also reported communication-related issues during maintenance and evolution that they associated with the presence of smells.

In this company, every component has a component owner, who is responsible of tracking changes, reviewing changes, handling questions from other owners or developers, as well as other organisational tasks.
This causes owners of components that are essentially god components to be overwhelmed with requests because their components implement a lot of functionality, have a lot of responsibilities and a lot of other components depend on them. 

Another problem in this category are code reviews of smelly components that contain a lot of files that change.
When the designers and architects meet to discuss and approve the changes in the code reviews, several discussions and arguments arise about the impact of each change, how to interpret the changes, and even what customers might be affected by certain changes thus creating confusion and ultimately delaying the development process.

\section{Discussion}\label{sec:discussion}
In this section we discuss the results obtained in this study and compare them to related work. Each subsection focuses on a significant aspect of the results we obtained from each research question. 

\subsection{Entanglement of dependencies}
An interesting observation stemming from the results obtained from RQ1 is that most of the cycles at file-level pose (in themselves) little threat to the maintainability level of the system as they (a) were not associated with bugs by our practitioners (unless they crossed the component border) and (b) only half of them survive for more than a year.
However, we noticed that when multiple cycles co-exist within the same component they create an  \textbf{entanglement of dependencies} that ultimately affects the clarity, testability, reusability, and the ability to anticipate the effects of changes of the parts affected by the cycles.
In fact, Lippert \cite{Lippert2006} hinted (back in 2006) at the possibility that cycles among files (or classes) may affect those aspects of Maintainability; the results presented in this paper corroborate his heuristics with empirical evidence.
Our results also align with those of Mo et al. that supported such heuristics in their industrial study \cite{Mo2018}.
More specifically, they found that clique-shaped cycles among files generated a considerable amount of maintenance activities in the affected components. 
The high coupling created by the presence of several cycles among the same group of files (such as cliques, or quasi-cliques) increases the \textbf{maintenance effort} required to maintain them.

On a similar note, Lippert had also mentioned that, while spaghetti code (i.e. \emph{goto} statements) is thought to be a thing of the past, modern software code still presents similar structures; but, instead of occurring at function or statement level, it involves files and components.
In other words, we \textbf{never really got rid of spaghetti code}'s negative effects (confusion, difficulty applying changes, intertwined logic etc.); we just solved the most explicit part of the problem, the one showing up in the code (i.e. the goto statements). 
Now we are facing the part of the problem that affects the way we organise code (files and components), where the negative effects can potentially have a larger impact.
The findings of this study show exactly this particular aspect, highlighting how practitioners struggle with maintaining entangled files and components and \textbf{need assistance} to manage the intricate structures that arise in their codebase.
Therefore, we advise researchers, to focus more on building tools and frameworks that reduce the burden of dealing with this particular type of issues, as well as on making these means more \emph{accessible and usable} by the industry.
While tools like \textsc{Arcan} are a first step towards this goal, the findings of this paper can  guide research activities in this direction too.
One example stems from our results on the introduction order of architectural smells. A machine learning tool that precisely predicts the introduction of new architectural smells in a component could be of great value to practitioners.

\subsection{Persistence of smells}
The results of RQ1.2 show that 50\% of Cyclic Dependencies do not survive more than 10 versions after their appearance.
Bavota et al. \cite{Bavota2015} studied the relationship between refactorings and code smells, and, surprisingly, their findings show that only 7\% of code smells are removed because of \emph{intentional} and specific refactoring activities. 
Should this finding be valid for architectural smells too, it would mean that only a \textbf{small percentage of architectural smells are intentionally removed by applying refactorings}.
The remainder of architectural smells may be therefore removed as part of the development activities related to the evolution of the system.
In our previous study, we also found that architectural smells' density over time is mostly constant in the long-term, meaning that as AS are removed from the system, they are also eventually replaced by others.
Cedrim et al.'s study \cite{Cedrim2017} report a similar percentage of code smells (i.e. 9.7\%) removed by refactorings, and, more interestingly, 33.3\% of refactorings actually resulted in the introduction of new code smells (most of which were never removed from the code).
Given our results, we could hypothesize that the same phenomenon may also occur for architectural smells: \emph{targeted refactorings potentially account for the minority of the architectural smells removed over time in a system}.
A possible explanation is that given the fact that architectural smells are not easy to visualise without proper tooling, then it is hard for developers and architects to realize what problem they are facing and thus act accordingly.

\subsection{Comparison with Java OSS}
Comparing the results obtained in RQ1 of this study with the results obtained in our previous study on Java OSS \cite{Sas2019}, we note both similarities and differences.

\paragraph{Evolution of smells}
In both cases, we found that the size of the smells either stays constant in size or increases over time while the smell density of the system remains constant.
Specifically, the size of the analysed systems (both C/C++ and Java) grows over time which entails that AS grow both in number and in size over time; this holds for both industrial C/C++ and Java OSS.
This is an expected result because software systems are expected to: (1) continue to grow over time (more lines of code are added every day); and (2) increase in complexity over time (more smells are added every day and existing smells may increase in size) \cite{Lehman1980}.

One difference we observed was that for OSS projects, UD had a dominant decreasing trend for its PageRank characteristic \cite{Sas2019}; this was not the case for C/C++ systems.
It is hard to objectively interpret this disparity given the different programming languages. However, we conjecture that the open source community is more successful in driving the more unstable components away from the centre of the system, where the maximally stable, core abstract components should reside \cite{Martin2017} and away from the implementation provided by the external ones.
It is important to note that the majority of Java OSS projects present in our previous study \cite{Sas2019} were Apache projects, which are known to follow high software quality standards.

\paragraph{Survivability of smells}
We noticed several similarities between Java OSS and the industrial C/C++ projects we analysed.
Both exhibit a trend where UD smells are the most persistent type of smell across the projects analysed.
HL smells follow UD in second place, which in turn are followed by CD smells (GC smells were not included in our original study).
Additionally, HL smells among components are more persistent than HL smells among files and cycles among components were less persistent than cycles among files in both types of systems.
Given these similarities, we can conclude that different architectural smell types exhibit the same pattern of persistence regardless of the type of system they are detected in.

However, we also noticed one important difference between the smells detected in Java OSS and C/C++ industrial systems: all \emph{smell types exhibit longer lifespans in the industrial systems}. 
This aligns with the feedback collected from our interviews with ASML engineers (see Section \ref{sec:results-rq-4-and-5}): making changes is hard, they require a lot of coordination between teams, certifications, code reviews, and a lot of effort in general.
The way ASML defines dependencies among components may also have had an impact on the survivability of smells. ASML components use a custom mechanism to expose their interface to other components. Thus, if a component communicates with another component through that interface, it is very likely that the dependency between those two components is there by design and not accidental or involuntary. This implies that it is less likely for that dependency to be removed in the future, and thus all smells relying on that dependency (e.g. like a cycle) will keep existing.
Ultimately, this custom mechanism, and the rigorous engineering processes between and within teams translate into an increased amount of time necessary to make a complete change to the system, which is also reflected in our data.

Finally, cycle shapes also exhibit the same patterns identified for Java OSS systems.
We especially observed that tiny cycles are outliving all other shapes in both types of systems, showing how this shape is very likely to be intentional and/or less harmful than the other types of cycle shape.

\subsection{Overlaps}
The results obtained from RQ2 show that, except for a few outliers, all smell types are \emph{likely} to overlap, \textbf{amplifying their impact on maintainability and evolvability} and \textbf{giving components more than one reason to change}, thus breaking the Single Responsibility Principle (SRP) \cite{Martin2017}. 

Our conversations with practitioners from RQ5 provide evidence to support this very claim, as they mentioned multiple examples where they associated two or more smells with the maintenance issues they were experiencing.
These results emphasise the importance of handling overlaps between architectural smells and, more importantly, preventing their introduction in the first place.

From our quantitative analysis emerged that cycles are pervasive in the system and they tend to appear as precursors to other smell instances, as they exhibit a high precedence rate (with $k = 1$).
This could mean that the presence of cycles in the system is likely to ease the introduction of other smells.
As a result, other smell types tend to have a high overlap rate (from 52\% to 77\% of instances, depending on the smell type) with cycles.
On the other hand, HL instances exhibit the opposite behaviour and have a low precedence rate but a rather high overlap with other smell instances (59\% to 77\% of HL instances).
This gives us an insight about the interplay between architectural smell instances of different types.
Cycles act as \emph{catalysts for more complex structures}, such as HL, to arise and negatively affect the maintenance of the affected components and files.
There were plenty of occasions where we observed star-shaped (see Figure \ref{fig:cycle-shapes}) cycle instances of which central element was also affected by a HL instance.
Indeed, in our RQ3 results, one of the drawbacks of a Hublike Dependency smell is that it aggregates responsibilities that it either delegates or implements itself.
Which just by itself breaks the SRP principle and impacts negatively maintainability.
For UD instances, on the other hand, tightly coupled structures such as cycles have an inherently high instability \cite{martin1994}, which in turn reflects to the component depending on them, thus creating an UD instance.
Understanding how and why CD instances are precursors to other smell instances is an interesting opportunity for future work.

\subsection{Feedback from practitioners}
From the results of RQ4, we found that AS analysis is quite useful to practitioners, especially for \emph{monitoring} purposes, rather than \emph{identification}: practitioners are mostly aware of the hotspots in their systems but they do need assistance in \emph{tracking} and \emph{quantifying} their presence deterministically.
Interestingly, these findings match what Mo et al. \cite{Mo2018} encountered but partially contrast the findings of Martini et al. \cite{Martini2018}, as in their case, practitioners were mostly unaware of the architectural smells in their system but found the information provided by smells still useful.

A possible explanation for this discrepancy is the fact that most of our participants have a long experience working for the company: they worked as developers for a long time in a project before becoming architects (or senior developers) of the same project. This means that they have a much more in-depth understanding of the problems in their system, so the information provided by a tool can mostly confirm this understanding.
We can only conclude that the level of awareness of developers of the smells in their system varies from subject to subject. In fact, a recent study on the topic \cite{Fontana2020} showed that developers' awareness of smells ranges from 26\% of all the smells detected in the system up to 78\%, depending on the participant.
This has a clear implication for researchers: if they are able to show the same information that a senior developer (or architect) is already familiar with, to all members of the team, regardless of their experience, then architectural smell analysis does \textbf{provide an added value to the team}.

Another common finding with Mo et al.'s work \cite{Mo2018} is the feedback of practitioners concerning the created reports.
Similarly to our study, Mo et al. also prepared reports that summarized the results to the developers and architects of the system.
These were very much appreciated by the developers and engineers in the companies of the two studies, and as a result, both companies showed interest in creating an integration with their own CI/CD to automate the analyses and provide daily (or weekly) reports.

One finding that was not reported by the subjects interviewed by Mo et al. is that our practitioners also highlighted the usefulness of our reports to new team members, and how they allow an easy transfer of knowledge to the less experienced members.

\subsection{Applying changes to the codebase}
The results of RQ5 show that practitioners struggle to maintain the components affected by architectural smells in a sustainable way. The main reasons include change propagation and the effects of a change in unknown parts of the codebase, during both typical maintenance (i.e. bug fixing, adaptations to new technologies) and evolution (addition of new features) tasks.
Previous studies from the literature corroborate these findings with data extracted by mining software repositories. Le et al. \cite{Le2018} found evidence that in open source Java systems the presence of architectural smells correlates with change-prone artefacts. Similar findings were also reported by Oyetoyan et al. \cite{Oyetoyan2015} on circular dependencies specifically.

The study of Vaucher et al. \cite{Vaucher2009} looked at the change proneness of God Classes, and showed that some God Classes are significantly less change-prone because they exist by design. 
While their findings refer to a different type of artefact (i.e. a code smell that is similar to GC but not exactly the same), they offer an insight on why some of the subjects we interviewed dismissed God Components as less detrimental (than other GCs). Specifically, God Components that are made by design are more easily understood by practitioners, because they understand their design and are thus better able to handle their complexity. 

We can thus conclude that change-prone artefacts and architectural smells are \emph{highly correlated} as this relation has been identified both quantitatively and qualitatively as well as independently by different studies.
This \emph{strengthens} the evidence about the increased effort required to maintain artefacts affected by architectural smells and \emph{highlights} the importance for practitioners to manage architectural smells.

Mo et al. also report about the experiences of developers when dealing with ripple effects \cite{Mo2018}. For instance, Mo et al. report on how developers consider the risk of performing a change to a file and that sometimes this is underestimated. This is corroborated by our findings. However, we also provide extra information about the ramifications caused by changes both at a company level (impacting several other teams) but also about the shortcuts that developers take in order to avoid the risks imposed by those changes.
As an example for the latter case, some developers admitted to intentionally duplicating entire files in order to avoid impacting other files with their changes.

\section{Implications for practitioners}\label{sec:implications}
Our results can help software engineers and architects to become aware of the side-effects associated with the presence of architectural smells within a large embedded systems company such as ASML.
Particularly, a few key points that practitioners should consider are the following:
\begin{itemize}
    \item the importance of \textbf{continuously monitoring} the presence of cycles among components/packages, as stated by the Acyclic Dependencies Principle \cite{Lippert2006}. Practitioners should especially oversee the components (or packages) that exhibit an excessive amount of internal cycles, as these may \emph{severely degrade} the overall maintainability of the component. On top of that, we also found that cycles are catalysts for other smells to arise;
    \item the \textbf{appearance of a Hublike Dependency} may be a clear signal that the affected part requires some refactorings, given that, as we found, this type of smell is likely to appear after other smells already affect a component.
    Cedrim et al. \cite{Cedrim2017} found that the most effective refactorings are the ones that target aggregator-like smells (such as Hublike Dependency and God Component), therefore this is a clear actionable point for practitioners;
    \item the \textbf{experiences shared by ASML} engineers can provide insightful details to other practitioners to avoid incurring similar issues such as change ripple effects, architecture erosion, and communication bottlenecks. To this end, practitioners should stick to architectural principles \cite{Martin2017} and guidelines and avoid the presence of severe architectural smell instances;
    \item the \textbf{integration of historical change-related information} of the components into decision-making processes through dashboards and reports. Our findings show that recurring changes are often associated with the presence of an architectural smell.
    Repairing artefacts that are commonly subject to maintenance work may ease the extra burden required to implement new features or fix bugs on the long-term.
\end{itemize}

\section{Threats to validity}\label{sec:threats-to-validity}
We identified the potential threats to validity for this study and categorised them using the classification proposed by Runeson et al. \cite{Runeson2012}: \emph{construct validity}, \emph{external validity}, and \emph{reliability}.
Internal validity was not considered as we did not examine causal relations \cite{Runeson2012}.

\paragraph{Construct validity}
This aspect of validity reflects to what extent this study measures what it is claiming to be measuring \cite{Runeson2012}.
To ensure we measure how AS evolve and how practitioners experience AS, we developed a case study using a well-known protocol template \cite{Brereton2008} that was reviewed by the first two authors and an external researcher in several iterations to ensure that the data to be collected would indeed be relevant to the research questions.

A possible threat to construct validity is the correctness of the parsing algorithm for the proprietary parts of the C/C++ compiler adopted by the company of this study. 
To mitigate this threat, we manually validated the parsing algorithm with a list of well-known files and components that had dependencies defined by proprietary constructs in the code.
The whole process was also supervised by one of the architects taking part in the study.

Another threat concerns the detection of the smells considered in this project, which depend on the implementation offered by \textsc{Arcan}.
Lefever et al. \cite{Lefever2021} have shown that different tools for technical debt measurement (including DV8, CAST, and SonarQube, but \emph{not} \textsc{Arcan}) have divergent, if not conflicting, results regarding which files are problematic in a system.
This is due to the fact that different tools make different assumptions, use different definitions of a smell, and have different implementations of how to detect a smell \cite{Lefever2021}.
Therefore, we can only state that our quantitative results obtained through \textsc{Arcan} may not be fully comparable with the results obtained by other tools.
However, this would be the case \emph{even if we used any other tool}, as shown by Lefever et al. \cite{Lefever2021}. 
Having said that, this threat can be considered \emph{partially mitigated}, as the definitions of each architectural smell used by \textsc{Arcan} are based on independent, previous work.
In particular, CD is based on the Acyclic Dependencies Principles \cite{Martin2017,Lippert2006}, HL and UD on the definitions provided by \cite{Samarthyam2016,martin1994}, and GC on Lippert's definition \cite{Lippert2006} (further improved upon by the authors of \textsc{Arcan}).
This cannot be said for other tools available, as many of them are based on previous work of the very authors of the tools, and therefore may potentially be biased.
Moreover, to guarantee that the results obtained by the \textsc{Arcan} tool are indeed in line with the definitions provided by previous work, the tool was used and evaluated in a number of studies \cite{Fontana2016, Biaggi2018, Sas2019}.

Yet another threat concerns the methodology we used to select the subjects for the interviews. Instead of using a probabilistic approach (i.e. random sampling) to sample our subjects, we sampled them based on convenience and circumstance. 
This was mainly due to two factors. First, it was up to the architects of each team to approve the interviews with their engineers.
Second, we could not interview subjects from all of the projects we analysed, as not all project architects were willing to provide participants.
Nonetheless, we managed to interview a good number of subjects, with more than one person per project in almost all cases, different levels of seniority and a balance in roles. Therefore, we consider this threat as, at least partially, mitigated.

\paragraph{External validity}
This aspect of validity reflects to what extent the results obtained by this study are generalisable to similar contexts.

External validity is limited by the fact that we only analysed the projects belonging to a single company with its primary business focused on a single domain.
The threat is partially mitigated by the fact that our quantitative results corroborate previous findings from open source systems: even though we only studied architectural smells in one, specific context, the findings have good chances to be applicable to other contexts as well.
Our qualitative results, on the other hand, can be applicable to large companies that employ a similar development process such as the company subject of this study.

Another threat to the generalisation of our results is the fact that the studied projects are part of a software product line composed of several products (machines).
This poses the risk of limiting the applicability of our results to the specific context of companies that develop software product lines.
The risk arises because architects and engineers must take into account the reuse of their code in different products with different hardware configurations, which may not be the case for many other companies.
To mitigate this risk, we focused our attention on issues that can occur independently of the practices adopted to develop and manage the software assets of the company.
For instance, during our interviews, we obtained a few data points mentioning reusability-specific issues encountered by engineers, however, we opted to only include in our results those that are potentially applicable to other contexts in order to not limit external validity.

\paragraph{Reliability}
Reliability is the aspect of validity focusing on the degree to which the data collection and analysis depend on the researchers performing them.

While we cannot share our dataset for confidentiality reasons, we do, however, provide a replication package\footnote{Visit \url{https://doi.org/10.6084/m9.figshare.16884739.v1}.} containing a complete version of the study design of this study and a sample of the report we sent to our practitioners to allow researchers to reuse similar data visualisations in their future work.
Moreover, the tools used in this study are freely available online\footnote{\label{fn:tools}See \url{https://github.com/darius-sas/astracker} and \url{https://gitlab.com/essere.lab.public/arcan}.} to allow other researchers to assess the rigour of the study or replicate the results using a different set of projects.

Another threat to reliability is the bias towards the data introduced by the researcher performing the coding.
This threat was mitigated by having a second researcher inspect both the codes and the coding maps extracted during each round of coding. All the feedback received was then integrated and the subsequent coding sessions adopted the updated codes.
The analysis was also performed using well-established techniques already used in previous work on the same topic as well as also in different fields (e.g. survival analysis, in the biomedical sciences field).
Therefore, we consider this threat mitigated.

\section{Conclusion and Future Work}\label{sec:conclusion-fw}
In this paper, we presented the results of an empirical embedded multiple-case case study performed using both quantitative and qualitative data. 
The data was collected by statically analysing 280 releases (spanning almost 3 years) across 9 industrial projects and by interviewing 12 subjects responsible for developing and architecting the projects under consideration.

To collect the quantitative data, we used a tool called \textsc{Arcan} to mine architectural smells (and their characteristics) from the over 20 millions lines of code available to us.
We then used different techniques to study the evolution of the architectural smells and understand how they evolve over time, how long they persist within the system depending on their type, and how they overlap with each other.
The findings show that smells grow over time in size, and that most of the detected instances do not persist for more than 2-3 releases. 
Moreover, most smell types were found to have high percentages of overlap with other smell types, meaning that it is not uncommon for components to be susceptible to problems caused by multiple types of smells, as also highlighted by our subjects during the interviews.

Indeed, practitioners found that our results aligned with their intuitions of where the issues were located and commented that tooling that helps them manage AS could be quite useful to them.
During the interviews, practitioners also mentioned rather interesting experiences where they struggled maintaining components affected by architectural smells, thus providing evidence of the negative effects of AS on Maintainability.

In conclusion, this paper provides a much clearer, and backed by empirical evidence, view on the issues experienced by practitioners in the presence of AS.
Our future work, we will further study how individual smell instances affect the work of developers and what aspects make an instance more severe than another.

\section*{Acknowledgements}
We would like to thank the Center for Information Technology of the University of Groningen for their support and for providing access to the Peregrine high performance computing cluster.

Lastly, many thanks to ASML for agreeing to give us access to their codebase, allow to run our analyses, and interview their engineers.

\section*{Declarations}
\subsection*{Funding}
This work was supported by the European Union's Horizon 2020 research and innovation programme under grant agreement No. 780572 SDK4ED \\(https://sdk4ed.eu/).

\subsection*{Availability of data, material, and code}
The data collected in this paper is not available publicly as it belongs to ASML.
We do however provide a replication package at the following link \url{https://doi.org/10.6084/m9.figshare.16884739.v1} that contains the protocol we used to collect the data and an anonymised report that we showed practitioners.
The package also includes the R code we used for data analysis.
The two tools we used to carry out the analysis are also available online (see footnote \ref{fn:tools}), though they do not contain the modifications applied to support ASML's code.

\subsection*{Consent to participate}
All subjects of our interviews were asked whether they would agree taking part in the interview beforehand. Interviewees were also informed of the possibility that excerpts from the interviews may be used in a published version of the manuscript.

\begin{appendices}
    
\section{Interview guide}\label{appendix:interview-guide}
\subsection{Questions}
This section lists the questions for the steps listed in the outline.
Total duration of the interview: \textbf{35 minutes} max.

\paragraph{Introduction (3 min.)}
\begin{enumerate}
    \item Introduce yourself, your job, and your goal.
    \item Briefly mention what this interview is for and disclaimer on how their responses are used.
    \item Feel free to expand on any topic or anecdote.
\end{enumerate}

\paragraph{Background information (2 min.)}
\begin{enumerate}[resume]
    \item What is your current official position?
    \item What is your role in this project? (day-to-day tasks example)
    \item How many years of experience do you have in the current position and in total?
\end{enumerate}

\paragraph{Results presentation (8 min.)}
\begin{enumerate}[resume]
    \item Share screen.
    \item Explanation of Architectural Smells and negative effects on maintenance activities (concise and stick to literature)
    \item Explanation in detail of each section of the report by going through each table and plot.
\end{enumerate}

\paragraph{General (2 min.)}
\begin{enumerate}[resume]
    \item General insights emerging from the analysis
    \begin{enumerate}
        \item What are one negative and one positive aspects about the FC that emerged after inspecting the architecture analysis results?
    \end{enumerate}
\end{enumerate}

\paragraph{RQ4 (10 min.)}
\begin{enumerate}[resume]
    \item Importance of smell types and characteristics
    \begin{enumerate}
        \item What type of smell do you think is the most important in your case? Why?
        \item For each smell type we calculate specific metrics (called smell characteristics),
        which one, for each smell type, is of most interest for you? Why?
    \end{enumerate}
    \item Perceived and actual quality of the system
    \begin{enumerate}
        \item Did your perception of the quality of the components change after inspecting the results? How/Why not? If yes, can you make an example?
        \item Does the presence of smells confirm what you already knew?
        \item Do the smells affect parts of the system that you were expecting to have issues with? Why/How come?
        \item Are there any missing parts in our analysis?
    \end{enumerate}
\end{enumerate}

\paragraph{RQ5 (10 min.)}
\begin{enumerate}[resume]
    \item Impact of smells on Maintainability
    \begin{enumerate}
        \item What types of smell do you deem to be more detrimental for the Maintainability of the system?
        \item Can you give an example of an issue you experienced while maintaining a component affected by a smell?
        \item Do you think it was related with the presence of a smell?
        \item Would it be hard to fix these issues? What aspects of the smell make it hard to do so (size, affected elements, overlaps)?
    \end{enumerate}
    \item Impact of smells on Evolvability (e.g. how easy it is to implement new features)
    \begin{enumerate}
        \item How have these issues affected the Evolvability (implementation of new functionality) of the affected parts?
        \item (If not answered before) Do you remember any issue hindering the addition of new features to any of the components affected by smells?
        \item Have you been discussing specific obstacles for the implementation of new features?
    \end{enumerate}
    \item Possible remediation strategies
    \begin{enumerate}
        \item What would be a possible quality-improvement plan that you could implement based on the information acquired from this report? And what would help you implement it?
        \item Do the results help you prioritising the issues to fix? If yes, how/why? If no, what could help?
    \end{enumerate}
\end{enumerate}
\paragraph{Feedback (2 min.)}
\begin{enumerate}
    \item General Feedback on the results
    \begin{enumerate}
        \item What view provided you with the most valuable insights (Smell characteristics, Dependency graph, DSM)? Why?
    \end{enumerate}
\end{enumerate}

\end{appendices}

\bibliography{bibliography}
\bibliographystyle{apalike} 

\end{document}